\documentclass[useASM, referee]{mn2e}
\usepackage{subfigure}
\input{psfig.sty}
\usepackage{graphicx}
\begin{document}
\title{Deviation of light curves of gamma-ray burst pulses from standard forms due to the
curvature effect of spherical fireballs or uniform jets  }
\date{2005 January 8}
\pubyear{????} \volume{????} \pagerange{2} \onecolumn
\author[Qin and Lu]
       { Y.-P. Qin$^{1,2}$, R.-J. Lu$^{1,2,3,4}$\\
$^1$National Astronomical Observatories/Yunnan Observatory,
Chinese Academy of Sciences,\\P. O. Box 110, Kunming 650011, China\\
$^2$Physics Department, Guangxi University, Nanning, Guangxi 530004, P. R. China\\
$^3$The Graduate School of the Chinese Academy of Sciences\\
$^4$ E-mail: luruijing@126.com
 }
\pagestyle{plain}
\date{Accepted ????.
      Received ????;
      in original form 2005 April 18}
\pagerange{\pageref{firstpage}--\pageref{lastpage}} \pubyear{2005}
\maketitle \label{firstpage}
\begin{abstract}

As revealed previously, under the assumption that some pulses of
gamma-ray bursts are produced by shocks in spherical fireballs or
uniform jets of large opening angles, there exists a standard
decay form of the profile of pulses arising from very narrow or
suddenly dimming local (or intrinsic) pulses due to the
relativistic curvature effect (the Doppler effect over the
spherical shell surface). Profiles of pulses arising from other
local pulses were previously found to possess a reverse S-feature
deviation from the standard decay form. We show in this paper
that, in addition to the standard decay form shown in Qin et al.
(2004), there exists a marginal decay curve associated with a
local $\delta$ function pulse with a mono-color radiation. We
employ the sample of Kocevski et al. (2003) to check this
prediction and find that the phenomenon of the reverse S-feature
is common, when compared with both the standard decay form and the
marginal decay curve. We accordingly propose to take the marginal
decay curve (whose function is simple) as a criteria to check if
an observed pulse could be taken as a candidate suffered from the
curvature effect. We introduce two quantities $A_1$ and $A_2$ to
describe the mentioned deviations within and beyond the $FWHM$
position of the decay phase, respectively. The values of $A_1$ and
$A_2$ of pulses of the sample are calculated, and the result
suggests that for most of these pulses their corresponding local
pulses might contain a long decay time relative to the time scale
of the curvature effect.

\end{abstract}


\section{Introduction}

In spite of the temporal structure of gamma-ray bursts (GRBs)
varying drastically, it is generally believed that some
well-separated pulses represent the fundamental constituent of GRB
time profiles (light curves) and appear as asymmetric pulses with
a fast rise and an exponential decay (FRED)(see, e.g., Fishman et
al. 1994).

Due to the large energies and the short timescales involved, the
observed gamma-ray pulses are believed to be produced in a
relativistically expanding and collimated fireball. The observed
FRED structure was interpreted by the curvature effect as the
observed plasma moves relativistically towards us and appears to
be locally isotropic (e.g., Fenimore et al. 1996, Ryde \&
Petrosian 2002; Kocevski et al. 2003, hereafter Paper I). Several
investigations on modeling pulse profiles have previously been
made (e.g., Norris et al. 1996; Lee et al. 2000a, 2000b; Ryde et
al. 2000, 2002). Several flexible functions describing the
profiles of individual pulses based on empirical relations were
presented. As derived in details in Ryde et al. (2002), a FRED
pulse can be well described by equation (22) or (28) there. Using
this equation, they could characterize individual pulse shapes
created purely by relativistic curvature effects in the context of
the fireball model.

Qin (2002) derived in details the flux intensity based on the
model of highly symmetric expanding fireballs, where the Doppler
effect of the expanding fireball surface is the key factor to be
concerned. The formula is applicable to cases of relativistic,
sub-relativistic, and non-relativistic motions as no terms are
omitted in the corresponding derivation. With this formula, Qin
(2003) studied how emission and absorbtion lines are affect by the
effect. Recently, Qin et al. (2004, hereafter Paper II) rewrote
this formula in terms of the integral of the local emission time,
which is in some extent similar to that presented in Ryde \&
Petrosian (2002), where relation between the observed light curve
and the local emission intensity is clearly illustrated. Based on
this model, many characteristics of profiles of observed gamma-ray
burst pulses could be explained. Profiles of FRED pulse light
curves are mainly caused by the fireball radiating surface, where
emissions are affected by different Doppler factors and boostings
due to different angles to the line of sight, and they depend also
on the width and structure of local pulses as well as rest frame
radiation mechanisms. This allows us to explore how other factors
such as the width of local pulses affect the profile of the light
curve observed.

Revealed in Paper II, there exists only a slowly decaying phase in
the light curve associated with a local $\delta $ function pulse,
for which no rising phase can be seen. For a local pulse with a
certain width, the light curve observed would contain both the
rising and decaying parts. It was revealed that light curves
arising from very narrow local pulses and those arising from
suddenly dimming local pulses share the same form of profiles in
their decaying phase, which was called a standard decay form (see
Paper II). For a common local pulse for which the decaying time is
not short enough, the profile of the decay portion of the
resulting light curve would significantly deviate from the
standard form. It is interesting that, the deviation could be
characterized by the feature of a reverse ``S'' (see Paper II Fig.
5). We wonder if this indeed holds for FRED pulse GRBs. If it
holds, how can we describe this deviation? Motivated by this, we
explore quantitatively in this paper the deviation of light curves
of gamma-ray burst pulses from the so-called standard decay form.
A sample of FRED pulse sources will be studied.

The paper is organized as follows. In section 2, we analyse the
deviation of light curve pulses associated with gamma-ray burst
spherical fireballs or uniform jets, from the standard form and a
marginal curve. In section 3, we examine the deviation deduced
from a sample detected by the BATSE instrument on board the
Compton Gamma Ray Observatory. Discussion and conclusions are
presented in the last section.

\section{ Theoretical analysis}

As derived in details in Qin (2002) and Paper II, the expected
count rate of a fireball within frequency interval $[\nu_1,\nu_2]$
can be calculated by
\begin{equation}
C(\tau )=\frac{2\pi R_c^3\int_{\widetilde{\tau }_{\theta ,\min }}^{%
\widetilde{\tau }_{\theta ,\max }}\widetilde{I}(\tau _\theta
)(1+\beta \tau
_\theta )^2(1-\tau +\tau _\theta )d\tau _\theta \int_{\nu _1}^{\nu _2}\frac{%
g_{0,\nu }(\nu _{0,\theta })}\nu d\nu }{hcD^2\Gamma ^3(1-\beta
)^2(1+\frac{\beta}{1-\beta}\tau )^2},
\end{equation}
which is just equation (21) in Paper II. This formula was derived
under the assumption that the fireball expands isotropically with
a constant Lorentz factor $\Gamma>1$ and the radiation is
independent of direction. Present in the formula, $\tau _\theta$
is a dimensionless relative local time defined by $ \tau _\theta
\equiv c(t_\theta -t_c)/R_c$, where $t_\theta$ is the emission
time (in the observer frame), called local time, of photons
emitted from the concerned differential surface $ds_\theta $ of
the fireball ($\theta$ is the angle to the line of sight), $t_c$
is a constant which could be assigned to any values of $t_\theta$,
and $R_c$ is the radius of the fireball measured at
$t_\theta=t_c$. In equation (1), $\widetilde{I}(\tau _\theta )$
represents the development of the intensity magnitude in the
observer frame, and $g_{0,\nu }(\nu _{0,\theta })$ describes the
rest frame radiation mechanism, with $\nu _{0,\theta }$ being the
rest frame emission frequency which is related to the observation
frequency $\nu $ by the Doppler effect. Variable $\tau$ used in
the formula is a dimensionless relative time defined by $ \tau
\equiv [c(t-t_c)-D+R_c]/R_c$, where $D$ is the distance of the
fireball to the observer, and $t$ is the observation time. As
analyzed in Qin (2002) and Paper II, the relative time $\tau$ is
confined by $1-\cos \theta _{\min }+(1-\beta \cos \theta _{\min
})\tau _{\theta ,\min }\leq \tau \leq 1-\cos \theta _{\max
}+(1-\beta \cos \theta _{\max })\tau _{\theta ,\max }$, and the
integral limits $\widetilde{\tau }_{\theta ,\min }$ and
$\widetilde{\tau }_{\theta ,\max }$ are determined by $
\widetilde{\tau }_{\theta ,\min }=\max \{\tau _{\theta ,\min
},(\tau -1+\cos \theta _{\max })/(1-\beta \cos \theta _{\max })\}
$ and $ \widetilde{\tau }_{\theta ,\max }=\min \{\tau _{\theta
,\max },(\tau -1+\cos \theta _{\min })/(1-\beta \cos \theta _{\min
})\},$ respectively, where we assign $\theta _{\min
}\leq\theta\leq\theta _{\max }$ and $\tau _{\theta ,\min }\leq\tau
_{\theta }\leq\tau _{\theta ,\max }$.

A local $\delta $ function pulse, $\widetilde{I}(\tau _\theta
)=cI_0\delta (\tau _\theta -\tau _{\theta ,0})/R_c$ (where $I_0$
and $\tau _{\theta ,0}$ are constants), when inserting it into
(1), would produce an observed light curve of equation (35) in
Paper II, which is
\begin{equation}
C(\tau )=\frac{2\pi R_c^2I_0\int_{\nu _1}^{\nu _2}\frac{g_{0,\nu
}(\nu _{0,\theta })}\nu d\nu }{hD^2}C_0(\tau ),
\end{equation}
with
\begin{equation}
C_0(\tau )\equiv \frac{(1+\beta \tau _{\theta ,0})^2(1+\tau
_{\theta ,0}-\tau )}{\Gamma ^3(1-\beta
)^2(1+\frac{\beta}{1-\beta}\tau )^2}.
\end{equation}
Note that the range of $\tau $, within which the radiation of the
local $\delta$ function pulse over the concerned area is
observable, is $ 1-\cos \theta _{\min }+(1-\beta \cos \theta
_{\min })\tau _{\theta ,0}<\tau <1-\cos \theta _{\max }+(1-\beta
\cos \theta _{\max })\tau _{\theta ,0}$ (see Paper II). As a
product of the local $\delta $ function pulse, light curves
$C(\tau )$ and $C_0(\tau )$ reflect nothing but the pure curvature
effect. For $\Gamma \gg1$, the term $\beta\tau/(1-\beta)$ in
equation (3) could be written as $\beta\tau/(1-\beta)\simeq
(t-t_0)/(R_c/2\Gamma^2c)$, where $t_0$ is a constant. We find that
$R_c/2\Gamma^2c$ is exactly the time scale of the curvature effect
(see equation [5] in Paper I). (One can check that, in terms of
local time, this curvature effect time scale becomes $R_c/c$.)

In the case of the local $\delta $ function pulse, as shown in
Paper II equation (37), $\nu _{0,\theta }$ and $\nu$ are related
by $\nu _{0,\theta }=[1+\beta\tau/(1-\beta) ](1-\beta )\Gamma
\nu/(1+\beta \tau _{\theta ,0} ) $, from which one gets $d\nu
_{0,\theta }/\nu _{0,\theta }=d\nu /\nu$. Thus, when taking
$g_{0,\nu }(\nu _{0,\theta })$ as a $\delta$ function (i.e., when
considering a mono-color radiation) and when interval
$[\nu_1,\nu_2]$ is large enough, $C(\tau )$ would become $C_0(\tau
)$ (differing only by a factor). Thus, $C_0(\tau )$ represents the
light curve arising from a local $\delta $ function pulse and
associated with a mono-color radiation. As the radiation concerned
(the GRB spectrum) is not a mono-color one and must last an
interval of time, we call function $C_0(\tau )$ as a marginal
decay curve.

As mentioned above, a sudden dimming or a very narrow local pulse
could produce a standard decay form of light curves (see Paper
II). According to Fig. 5 of Paper II, the standard form could be
represented by equation (2). (Note that, a local $\delta$ function
pulse is an extremely narrow local pulse, and it belongs to the
class of suddenly dimming local pulses.) Shown in Preece et al.
(1998, 2000), $\alpha _0=-1$ and $\beta _0=-2.25$ are typical
values of the lower and higher indexes of the Band function
spectrum, deduced from most GRBs observed. We thus define $C(\tau
)$ associated with the rest frame Band function spectrum of
$\alpha _0=-1$ and $\beta _0=-2.25$ as a standard decay form which
was mentioned in Paper II previously.

With these two curves, we are able to explore how light curves
associated with different local pulse forms deviate from standard
ones, and with the deviation we might be able to estimate some
parameters of local pulses.

Taking $\tau _{\theta ,0}$=0, $\theta_{min}$=0, and
$\theta_{max}=\pi/2$, we get $0< \tau <1$, and equation (3)
becomes
\begin{equation}
C_0(\tau )=\frac{(1-\tau )}{\Gamma ^3(1-\beta
)^2(1+\frac{\beta}{1-\beta}\tau )^2}.
\end{equation}
For the sake of comparison, we normalize light curves of (2) and
(4) in intensity and re-scale its variable, $\tau$, by
$\tau=a\tau'+b$, so that the peak count rate is located at
$\tau'=0$ and the FWHM position of the decay portion is located at
$\tau'=0.2$ (see Paper II).

Formula (1) suggests that, except the state of the fireball (
i.e., $\Gamma$, $R_c$ and D), light curves of sources depend only
on $\widetilde{I}(\tau _\theta )$ and $g_{0,\nu }(\nu _{0,\theta
})$. We assume in this paper the common empirical radiation form
of GRBs as the rest frame radiation form, the so-called Band
function (Band et al. 1993) that could well describe spectra of
most sources (see, e.g., Schaefer et al. 1994; Ford et al. 1995;
Preece et al. 1998, 2000), and adopt different forms of local
pulses, $\widetilde{I}(\tau _\theta )$ in equation (1), which will
produce different light curves, to study the deviation from the
standard forms.

Let us consider two local pulses. The first is a local pulse with
an exponential rise
\begin{equation}
\widetilde{I}(\tau _\theta )=I_0\exp (\frac{\tau _\theta -\tau
_{\theta ,\max }}\sigma )\qquad \qquad \qquad (\tau _{\theta ,\min
}\leq \tau _\theta \leq \tau _{\theta ,\max }),
\end{equation}
and the second is a local pulse with an exponential decay
\begin{equation}
\widetilde{I}(\tau _\theta )=I_0\exp (-\frac{\tau _\theta -\tau
_{\theta ,\min }}\sigma )\qquad \qquad \qquad \qquad \qquad (\tau
_{\theta ,\min }\leq \tau _\theta ).
\end{equation}
Following Paper II, we take $\tau_{\theta,max}=10\sigma +
\tau_{\theta,min}$ (in
this case the interval between $%
\tau _{\theta ,max}$ and $\tau _{\theta ,\min }$ would be large
enough to make the rising part of the local pulse close to that of
the exponential pulse) and $\tau_{\theta,min}=0$. Light curves
arising from these local pulses are normalized and re-scaled in
the way mentioned above, which are shown in Fig. 1.
\begin{figure}
\centering
\includegraphics[width=5in,angle=0]{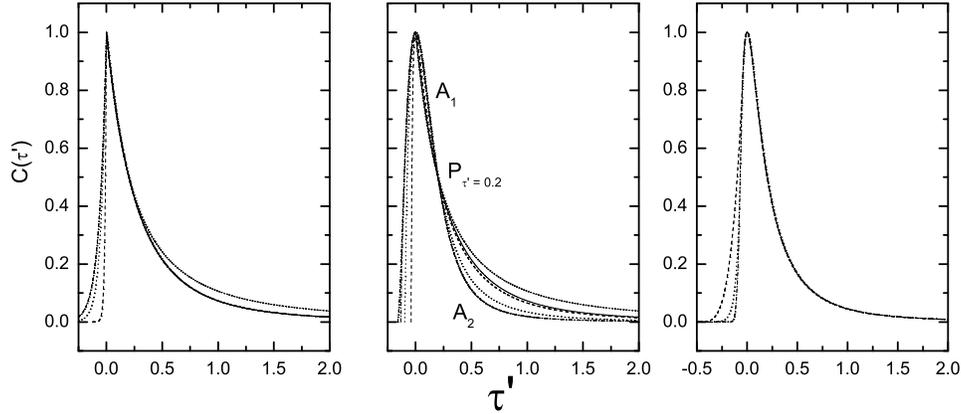}
\caption{ --Plots of the normalized and re-scaled light curve
$C(\tau')$ arising from the local exponential rise pulse (5) (the
left panel), the local exponential decay pulse (6) (the middle
panel), and three other local pulses (7) (dash line), (8) (dot
line), and (9) (dash dot line) (the right panel). The solid and
short dash lines represent the standard decay form and the
marginal decay curve in the left and middle panels, respectively.
A Band function rest frame radiation form with $\alpha _0=-1$ and
$\beta _0=-2.25$, and the frequency range of 100 $\leq$
$\nu/\nu_{0,p}$$ \leq$ 300, are adopted, and we take $\Gamma=100$,
$\sigma$=0.02 (dash line), 0.2 (dot line), 2 (dash dot line), and
20 (dash dot dot line), respectively. }
\end{figure}

We find from Fig. 1 that there is no significant deviation of the
light curves arising from local exponential rise pulses from the
standard decay form in the decay portion of the light curve, just
as what illustrated in Paper II. However, there are a significant
``positive'' deviation (over the standard form) of the light
curves associated with local exponential decay pulses from the
standard decay form within the range of $\tau' \in [0.0, 0.2]$,
and a significant ``negative'' deviation in the range of
$\tau'>0.2$, which we call a reverse ``S" deviation
characteristic. In addition, one finds from Fig. 1 that there
exists a ``negative'' deviation of the standard decay form from
the marginal decay curve within the range of $\tau'>0.2$, but
within the range of $\tau' \in [0.0, 0.2]$ there is no deviation
between the two.

We define the positive and negative deviation areas from the
standard decay form (and/or the marginal decay curve) as $A_1$ and
$A_2$, respectively. Relation between $A_1$ and $A_2$ and those
between these areas and $\sigma$ are presented in Fig. 2 (here
$A_2$ is calculated within the range of $\tau' \in [0.2, 2.0]$, as
light curves of local exponential decay pulses overlap each other
in the range of $\tau'>2$). As shown in Fig. 2, $log A_1$ is
linearly correlated with $log A_2$. In addition, we find that $log
A_1$ and $log A_2$ increase linearly with $log \sigma$ within the
range of $\sigma<0.05$. A linear analysis produces $logA_1 =
-0.993 + 0.873 log \sigma$ and $logA_2 = -0.278 + 0.952
log\sigma$. However, when $\sigma$ being large enough (say
$\sigma>2$), $A_1$ and $A_2$ would not change with $\sigma$ (in
other wrods, they are saturated).

To check if local pulses with different rising curves but sharing
the same decaying curve would lead to much different profiles of
light curves in the decay phase, we consider three other local
pulses. The first consists a power law rise and an exponential
decay:
\begin{equation}
\widetilde{I}(\tau _\theta )=I_0\{
\begin{array}{c}
(\frac{\tau _\theta -\tau _{\theta ,\min }}{\tau _{\theta ,0}-\tau
_{\theta ,\min }})^\mu \qquad \qquad \qquad \qquad \qquad (\tau
_{\theta ,\min }\leq\tau _\theta \leq \tau _{\theta ,0}) \\
\exp (-\frac{\tau _\theta -\tau _{\theta ,0}}\sigma )\qquad \qquad
\qquad \qquad \qquad (\tau _{\theta ,0}<\tau _\theta \leq \tau
_{\theta ,\max })
\end{array}.
\end{equation}
The second is an exponential rise and an exponential decay local
pulse:
\begin{equation}
\widetilde{I}(\tau _\theta )=I_0\{
\begin{array}{c}
\exp (\frac{\tau _\theta -\tau _{\theta ,0}}{\sigma_{1}} )\qquad
\qquad\qquad\qquad \qquad (\tau _{\theta ,\min }\leq \tau _\theta \leq \tau _{\theta ,0}) \\
\exp (-\frac{\tau _\theta -\tau _{\theta ,0}}\sigma )\qquad \qquad
\qquad \qquad \qquad (\tau _{\theta ,0}<\tau _\theta )
\end{array}.
\end{equation}
The third is a Gaussian rise and an exponential decay local pulse:
\begin{equation}
\widetilde{I}(\tau _\theta )=I_0\{
\begin{array}{c}
\exp [-(\frac{\tau _\theta -\tau _{\theta ,0}}{\sigma_{1}}
)^2]\qquad \qquad \qquad \qquad \qquad
(\tau _{\theta ,\min }\leq \tau _\theta \leq \tau _{\theta ,0}) \\
\exp (-\frac{\tau _\theta -\tau _{\theta ,0}}\sigma )\qquad \qquad
\qquad \qquad \qquad (\tau _{\theta ,0}<\tau _\theta )
\end{array}.
\end{equation}
We take $\sigma = 2.0$ for the decaying part of these three local
pulses. For the first one we take $\mu=2$ and $FWHM=2.0$, for the
second and third ones we adopt $\sigma_{1}=2.0$.

Light curves arising from these local pulses are presented in Fig.
1 as well. One finds that the three light curves possess the same
decaying profile. This suggests that the profile of the decaying
part of the light curve is determined only by the decaying curve
of the corresponding local pulses.

Besides the local pulses discussed above, we also study other
forms of local pulses, such as a power law rise and power law
decay pulse as well as a Gaussian local pulse. Relations between
$\sigma$ or $FWHM$ and $A_1$ and $A_2$ for the light curves
arising from these local pulses are presented in Fig. 3. The same
conclusions obtained above hold for these cases.

According to the analysis above, one finds that: a) light curves
associated with local pulses without a decaying portion would bear
the standard decay form in their decaying phase, which was
concluded previously in Paper II; b) there would be a reverse ``S"
feature deviation of the light curves arising from local pulses
containing a decaying portion, from the standard decay form, which
could also be concluded from Fig. 3 of Paper II; c) the deviation
concerned could quantitatively described by areas $A_1$ and $A_2$
defined above, and these two quantities are linearly correlated
with the width of the decaying curve of the local pulse when the
latter is small enough (say, when $\sigma<0.05$ in the case of an
exponential decaying local pulse); d) all curves associated with a
continuum spectrum (including the curve of the standard decay
form) are well below the marginal decay curve beyond the $FWHM$
position of the decay phase (say, within the range of
$\tau'>0.2$); e) within the $FWHM$ position of the decay phase
(say, in the range of $\tau' \in [0.0, 0.2]$), the standard decay
form and the marginal decay curve are not distinguishable.
\begin{figure}
\centering
\includegraphics[width=5in,angle=0]{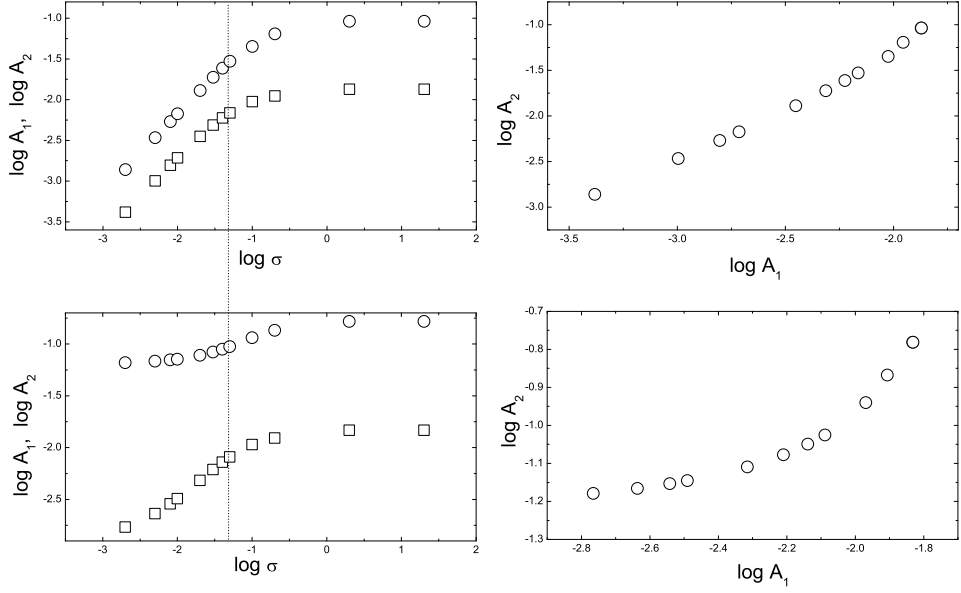}
\caption{ --Plots of $log A_1$ and $log A_2$ versus $\log \sigma$
(the left panels), and $log A_2$ versus $log A_1$ (the right
panels) for a typical local pulse. In the two left panels, the
open square and open circle represent $A_1$ and $A_2$,
respectively, associated with a local pulse with an exponential
rise and an exponential decay. Quantities $A_1$ and $A_2$
presented in the two upper panels represent the deviation of
pulses from the standard decay form, and those in the two lower
panels describe the deviation from the marginal decay curve. Other
parameters are the same as those adopted in Fig. 1. }
\end{figure}
\begin{figure}
\centering
\includegraphics[width=5in,angle=0]{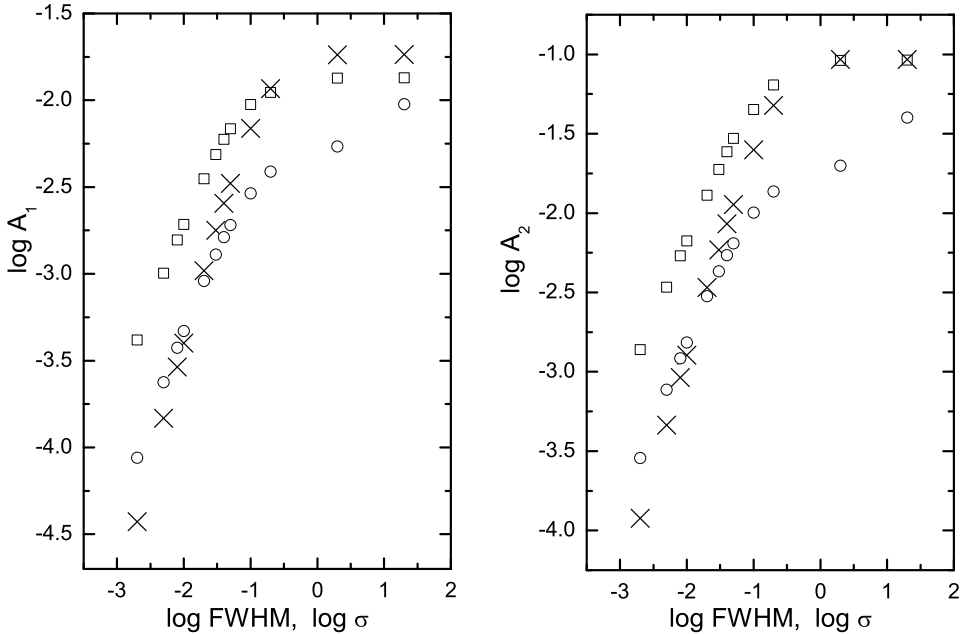}
\caption{ --Plots of $log A_1$ versus $log FWHM$ or $\log \sigma$
(the left panel), and $log A_2$ versus $log FWHM$ or $\log \sigma$
(the right panel) for some local pulses. Quantities $A_1$ and
$A_2$ represent the deviation of the light curves associated with
the local exponential decay pulse (the open square), the local
power law decay pulse (with the power law index $\mu$=2) (the
cross), and the local Gaussian pulse (the open circle), from the
standard decay form. Other parameters are the same as those
adopted in Fig. 1.}
\end{figure}

\section{Decaying form seen in a FRED pulse sample}

At least two questions urge us to employ a FRED pulse sample to
make the following analysis. One is whether or not the reverse
S-feature deviation characteristic indeed exists in the observed
light curves, and the other is if the profiles of these light
curves are indeed well below the marginal decay curve beyond the
$FWHM$ position of the decay phase.

To study this issue, we utilize the light curves of the sample
presented in Paper I, where the data are provided by the BATSE
instrument on board the CGRO (Compton Gamma Ray Observatory)
spacecraft. Of this sample, we find only the data of 75 individual
pulses, which are employed in the following. It is well-known that
pulses of a GRB show a tendency to self-similarity for each energy
band (see, e.g. Norris et al. 1996). We thus consider in this
paper only the light curve of channel 3 of BATSE, as signals in
this channel are always significant.

The background of light curves is fitted by a polynomial
expression using 1.024s resolution data that are available from 10
minutes before the trigger to several minutes after the burst. The
data along with the background fit coefficients can be obtained
from the CGRO Science Support Center (CGROSSC) at NASA Goddard
Space Flight Center through its public archives. All of the
background-subtracted light curves are fitted with equation (22)
of Paper I, and then we normalize and re-scale the data of the
background-subtracted light curves, using the method adopted
above, with the corresponding fitting curves (the magnitude and
$FWHM$ position of the background-subtracted light curves data
could be well estimated from these fitting curves). We find that
all pulses in our sample exhibit a reverse S-feature deviation
from the marginal decay curve, where the (central) profiles of the
light curves are indeed well below the marginal decay curve beyond
the $FWHM$ position of the decay phase (see Figs. 4 and 5). When
compared with the standard decay form, all of them except two,
$\#$ 3257 and $\#$ 5495, show the reverse S-feature deviation as
well (for the details see Fig. 5). A list of the deviation areas,
$A_1$ and $A_2$, from the marginal decay curve and the standard
decay form, are shown in Table 1, and their distributions are
presented in Fig. 6.

Comparing Fig. 6 with Fig. 3 we find that for most pulses of the
sample the widths of the decay phase of their corresponding local
pulses are sufficiently large (larger than 0.1) that they are no
more sensitive to the two areas $A_1$ and $A_2$. This suggests
that, compared to the curvature delay time of the emitting region
of the shock (see what discussed in last section), the intrinsic
pulse decay times are long. This is because the FRED shape due to
a local $\delta$-function pulse has a characteristic duration set
by the curvature time delay --- the time delay between the arrival
times of two simultaneously emitted photons, one from the
line-of-sight and one from $\theta_{max}$, but that due to a local
pulse with a sufficiently long decay time has a different
characteristic. At a certain observation time, only photons
emitted from a limited area could reach the observer in the case
of the local $\delta$-function pulse (in this case, the area could
be marked by $\theta$
--- $\theta + \Delta \theta$, when $\Delta \theta$ is extremely
small), while in the case of the long decay time local pulse,
photons emitted from all the areas concerned could be observed
(they are emitted from different local times)(note that, in the
case of the suddenly dimming local pulse, the corresponding area
would decline with time). It seems that it is this difference that
leads to the variance of the light curve characteristic seen in
the two cases. Under this interpretation, we suspect that the fact
that many BATSE pulses are in the saturation regime suggests that
constraints may be placed on the angular size and radius of the
emission region (which might deserve a further investigation).
\newpage
\begin{figure*}
\resizebox{5cm}{!}{\includegraphics{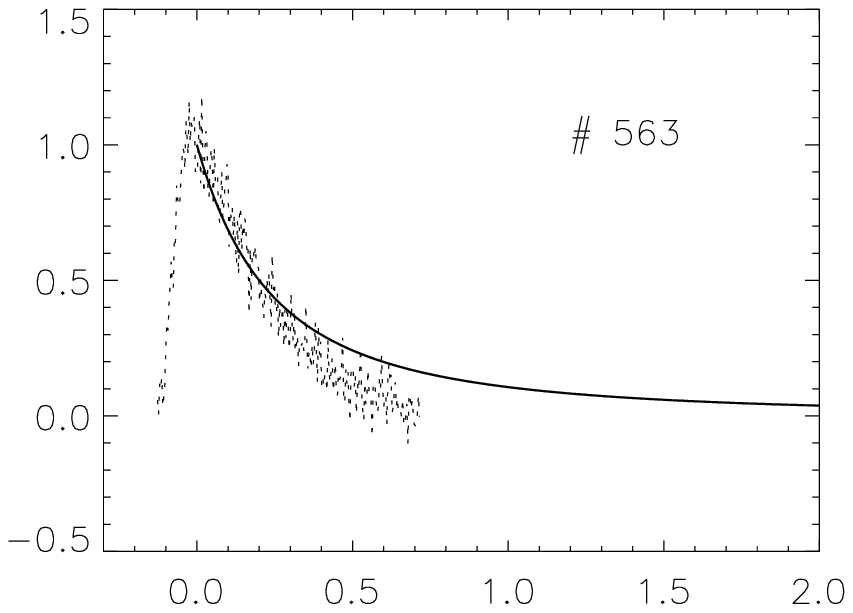}}
\resizebox{5cm}{!}{\includegraphics{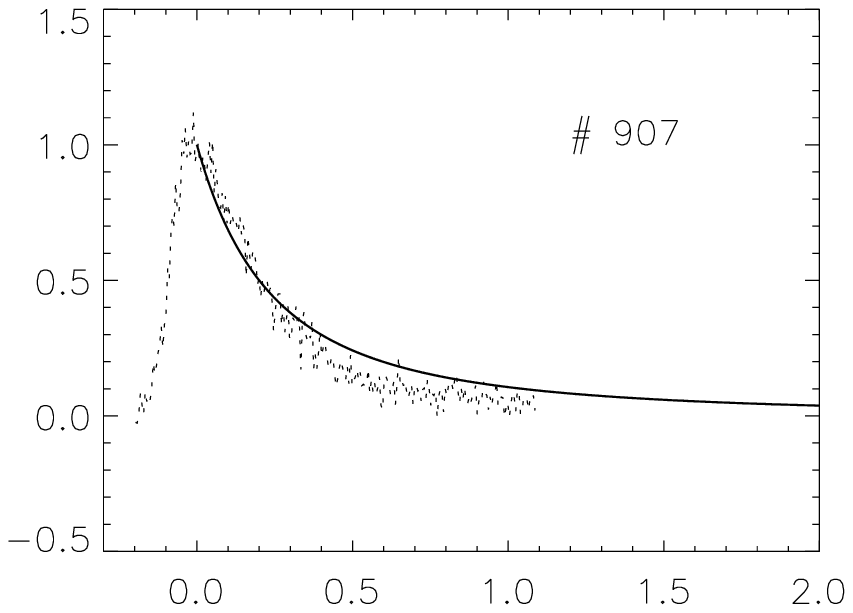}}
 \resizebox{5cm}{!}{\includegraphics{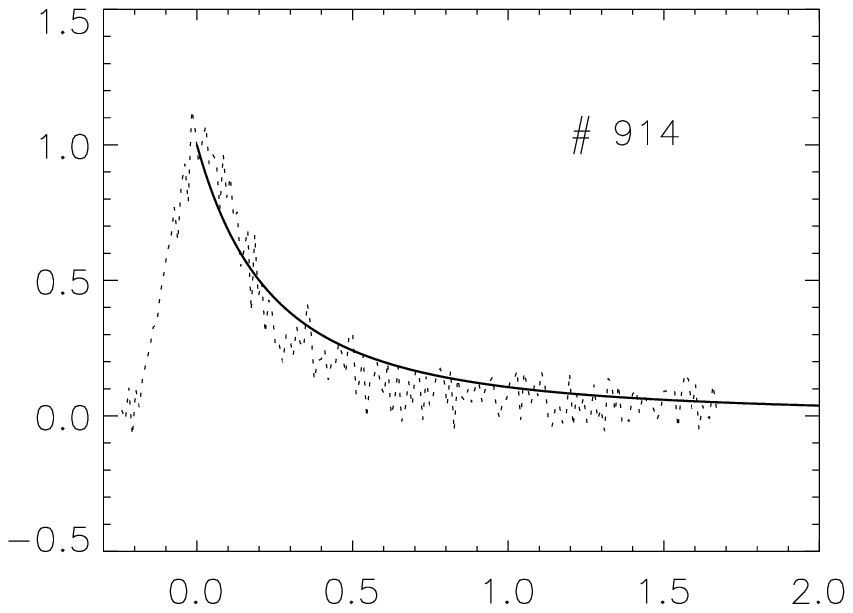}}
 \resizebox{5cm}{!}{\includegraphics{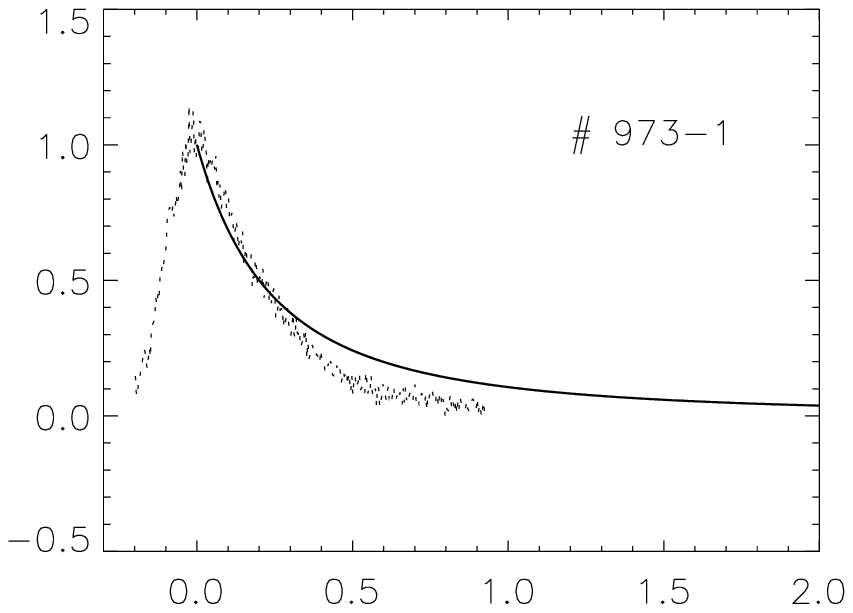}}
 \resizebox{5cm}{!}{\includegraphics{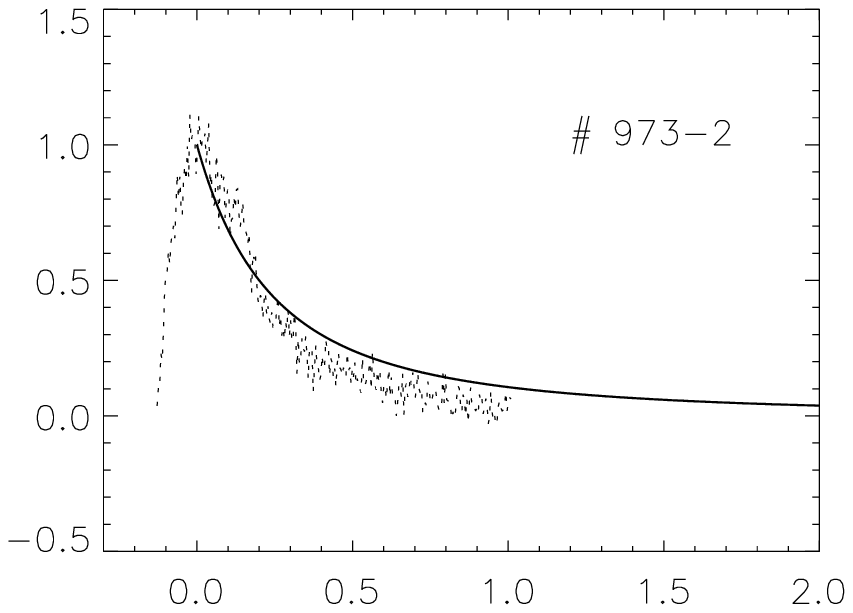}}
 \resizebox{5cm}{!}{\includegraphics{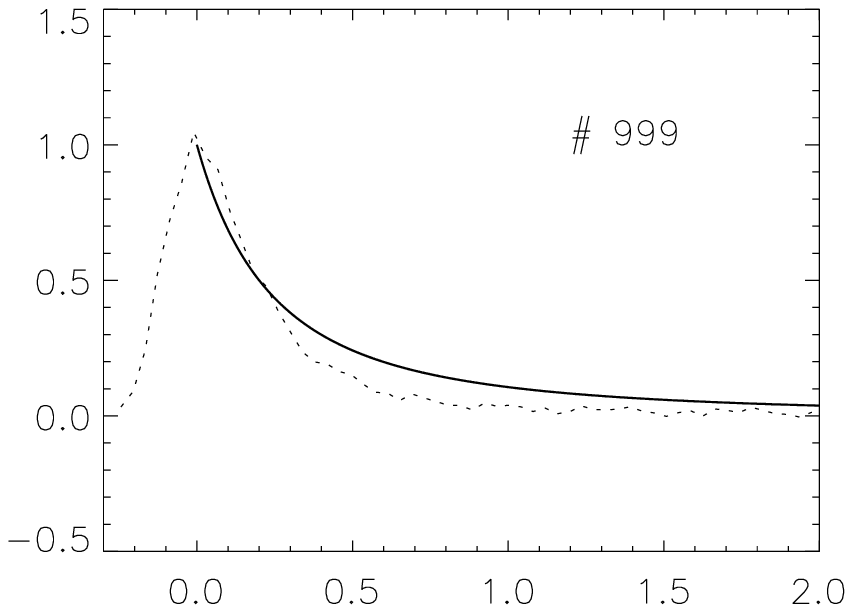}}
 \resizebox{5cm}{!}{\includegraphics{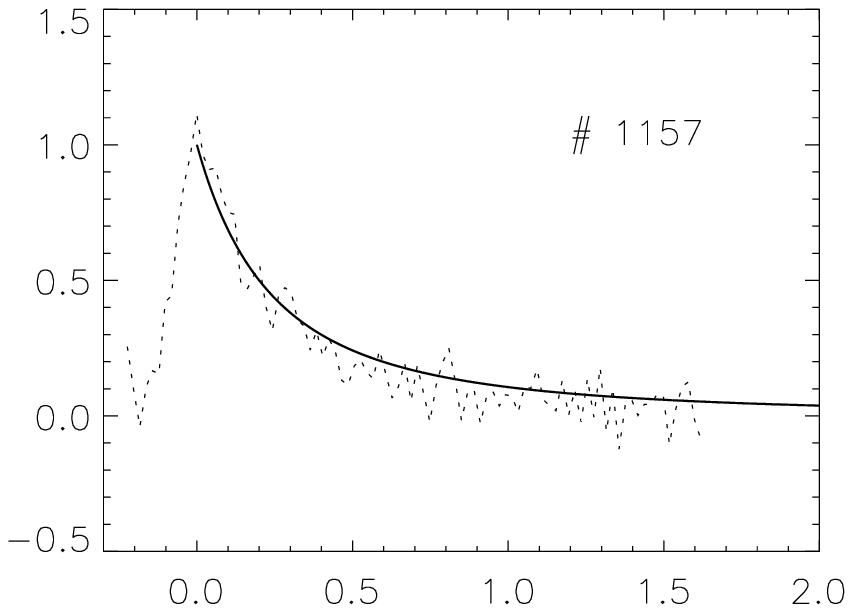}}
 \resizebox{5cm}{!}{\includegraphics{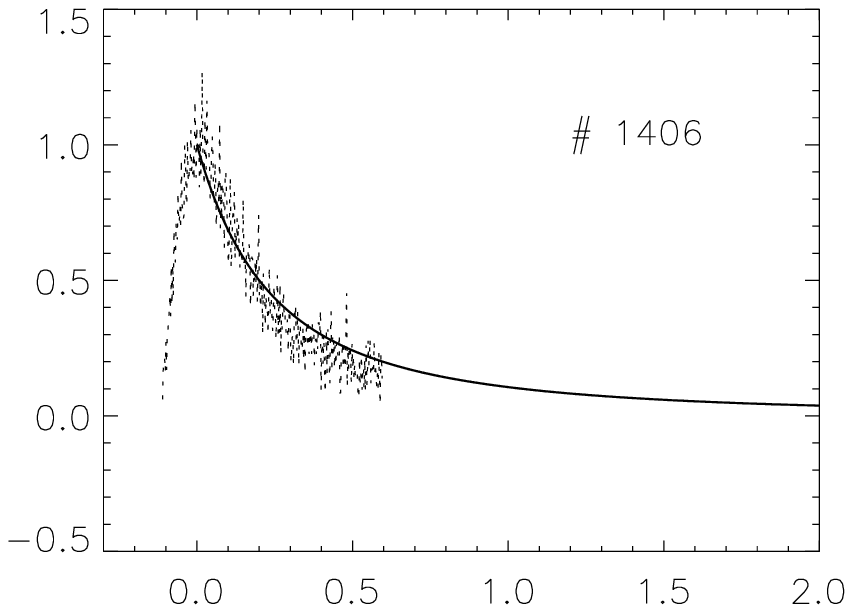}}
 \resizebox{5cm}{!}{\includegraphics{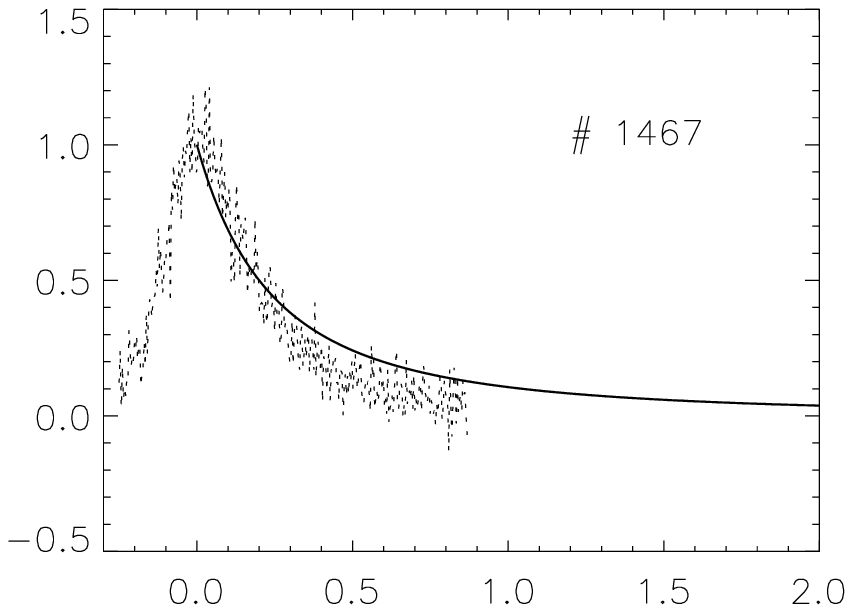}}
 \resizebox{5cm}{!}{\includegraphics{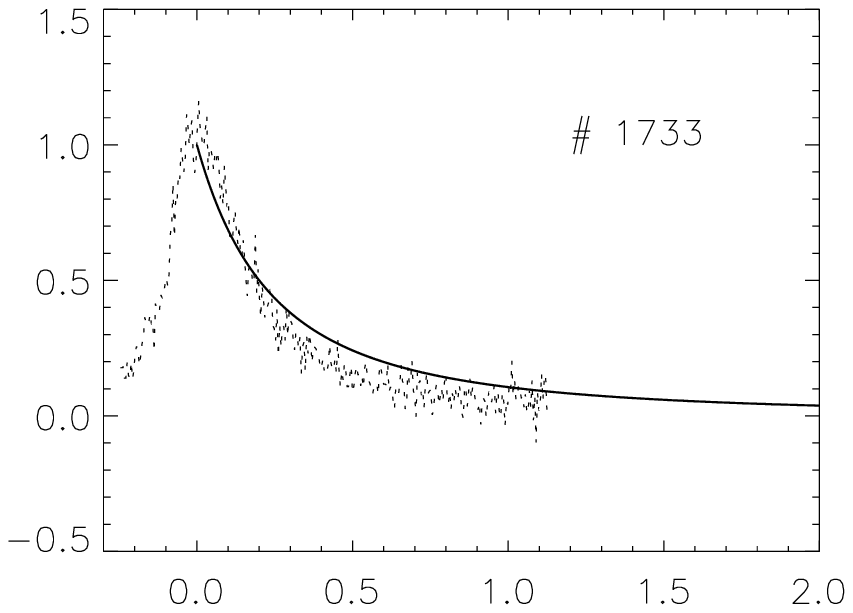}}
  \resizebox{5cm}{!}{\includegraphics{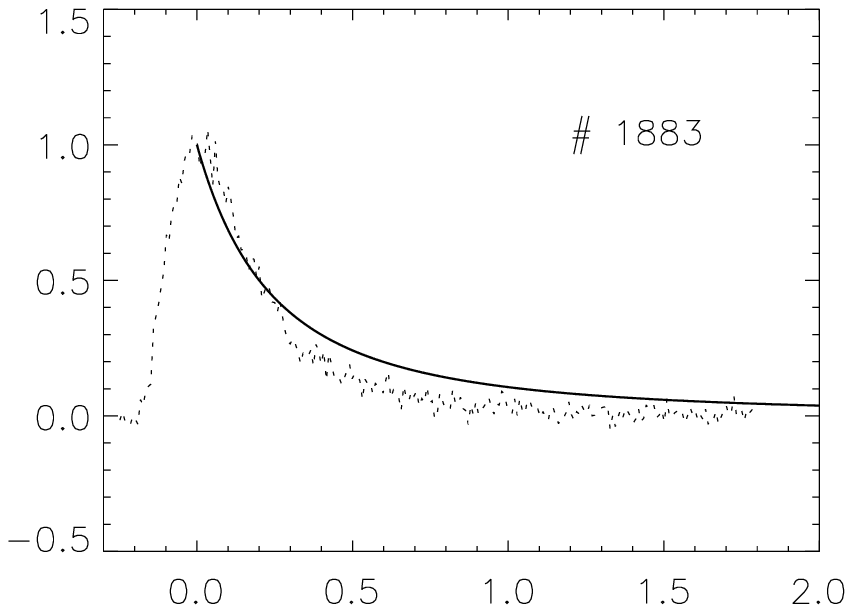}}
 \resizebox{5cm}{!}{\includegraphics{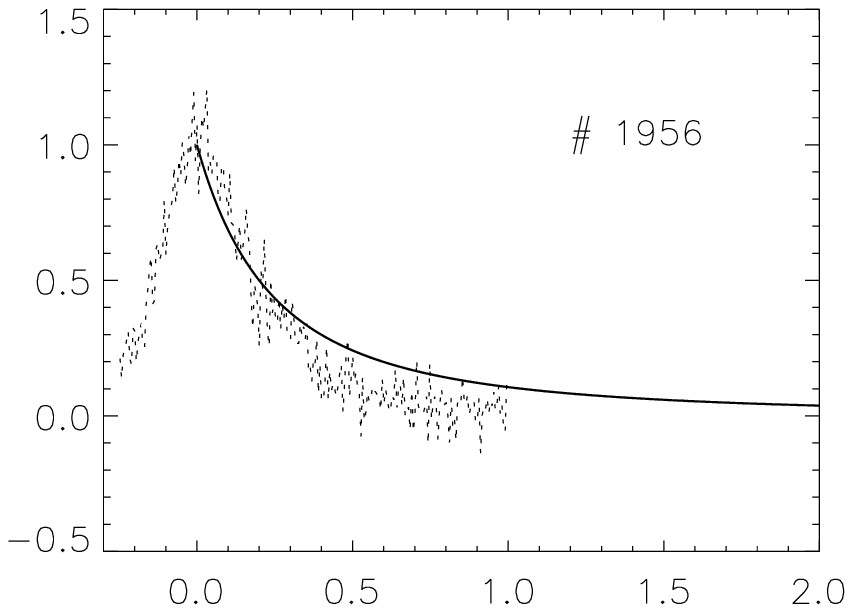}}
 \resizebox{5cm}{!}{\includegraphics{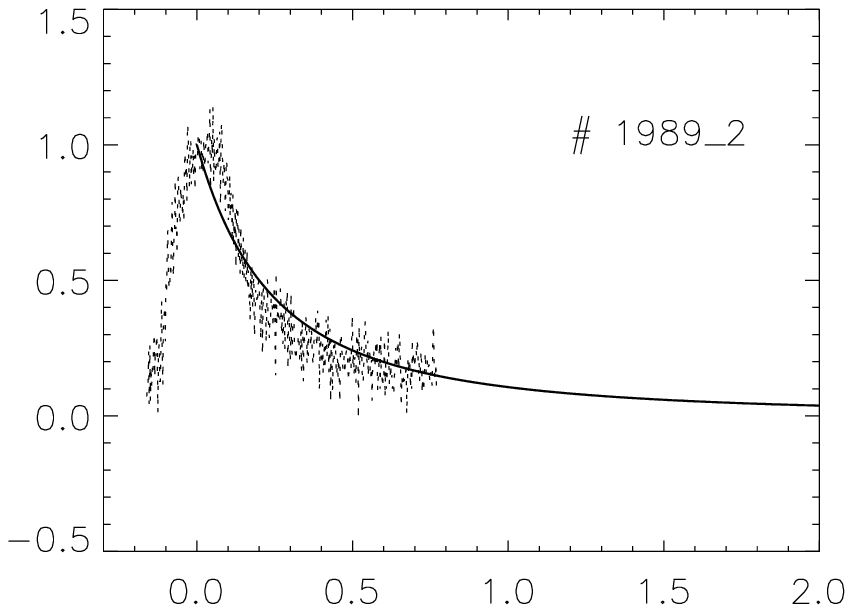}}
 \resizebox{5cm}{!}{\includegraphics{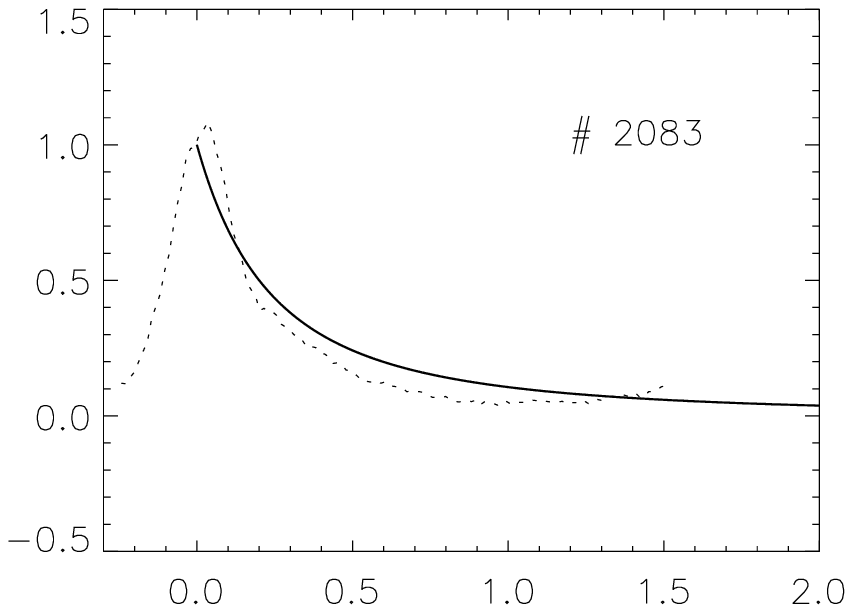}}
 \resizebox{5cm}{!}{\includegraphics{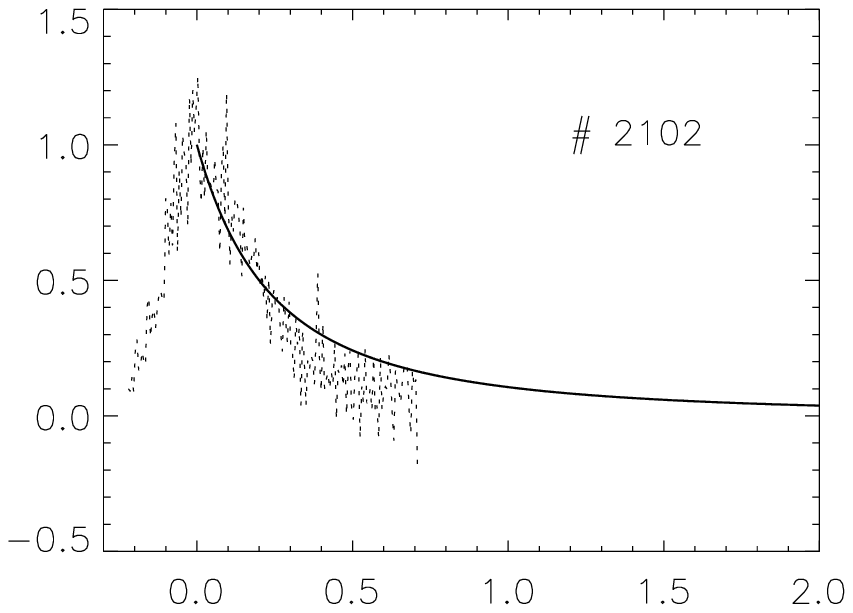}}
 \resizebox{5cm}{!}{\includegraphics{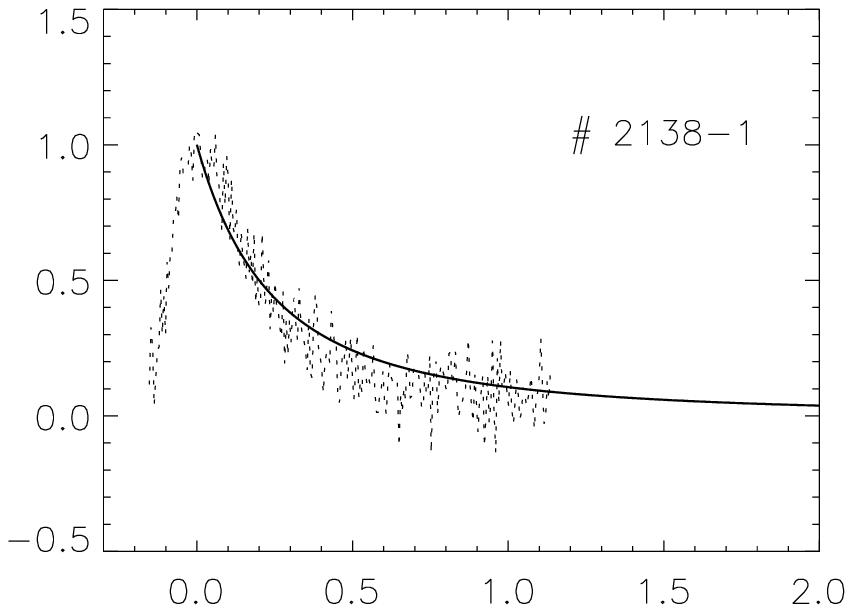}}
 \resizebox{5cm}{!}{\includegraphics{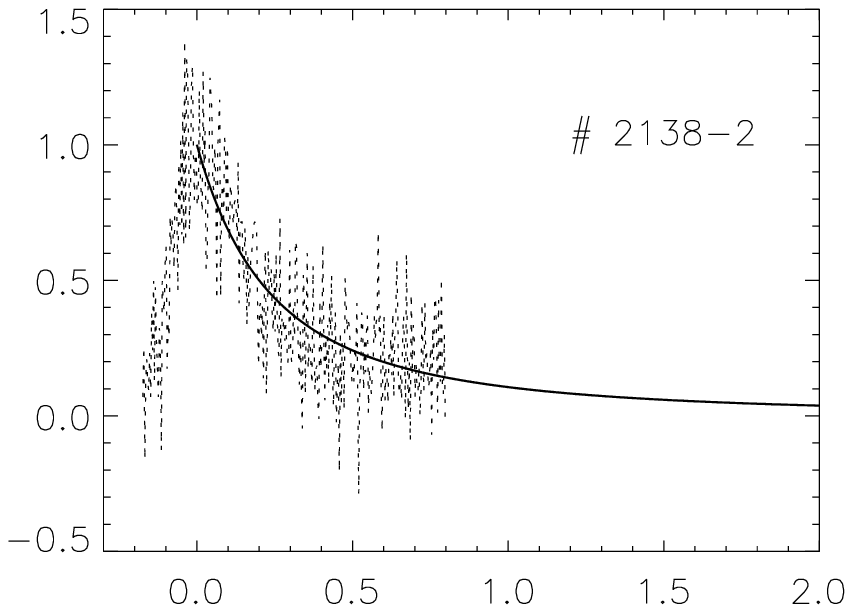}}
 \resizebox{5cm}{!}{\includegraphics{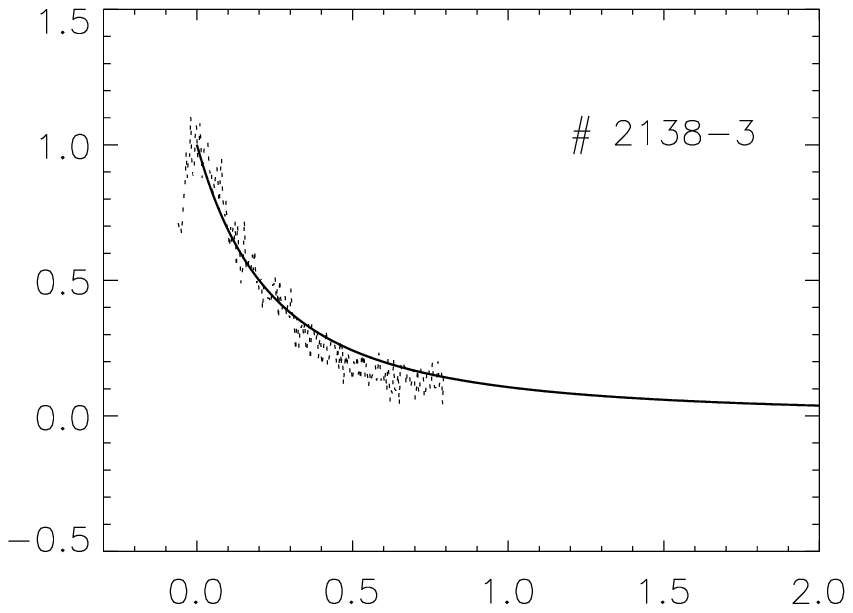}}

 \caption{--Plots of the observed
light curves (the dot line) and the marginal decay curve (the
solid line) for the 75 individual pulses of the sample.}
 \label{sed}
\end{figure*}

\newpage
\begin{figure*}
\resizebox{5cm}{!}{\includegraphics{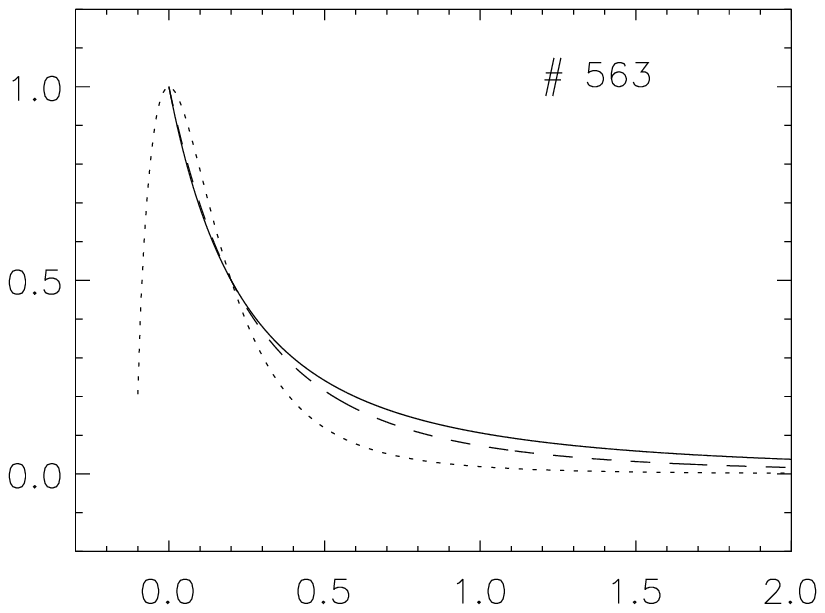}}
\resizebox{5cm}{!}{\includegraphics{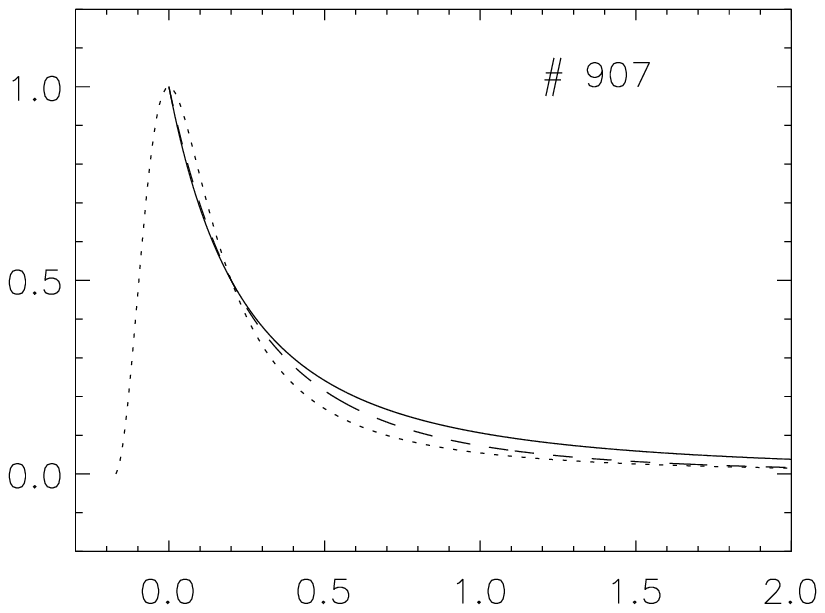}}
 \resizebox{5cm}{!}{\includegraphics{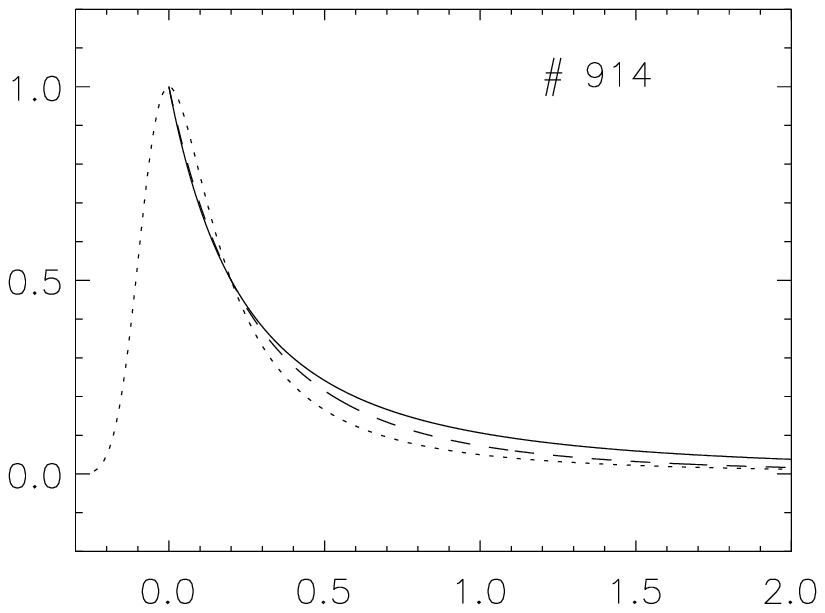}}
 \resizebox{5cm}{!}{\includegraphics{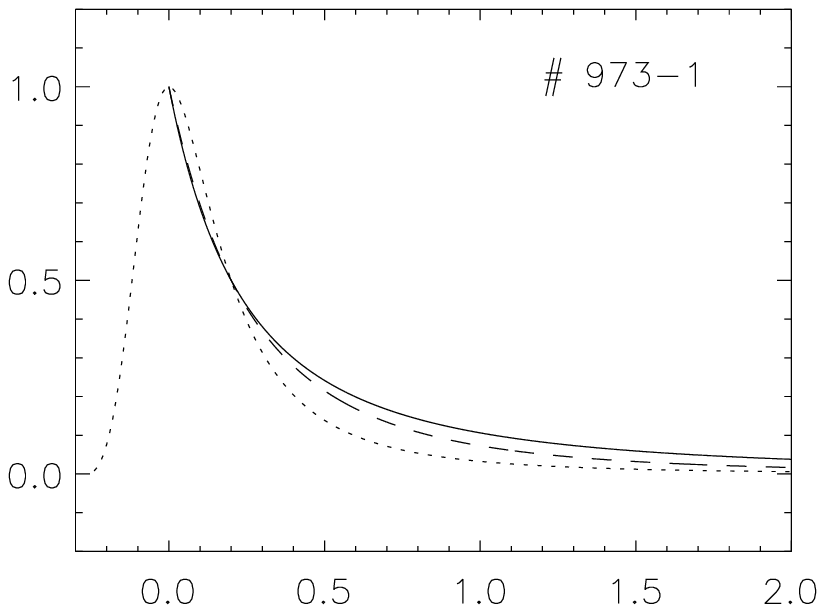}}
 \resizebox{5cm}{!}{\includegraphics{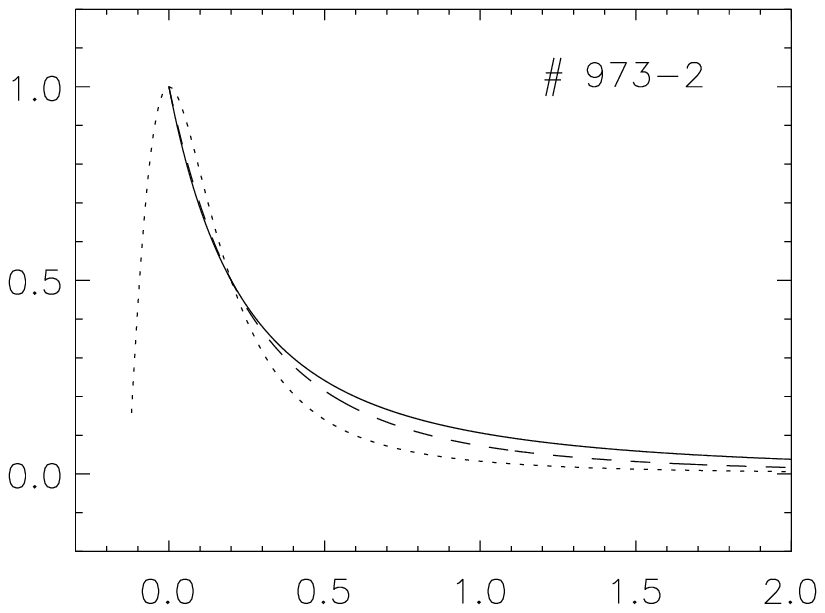}}
 \resizebox{5cm}{!}{\includegraphics{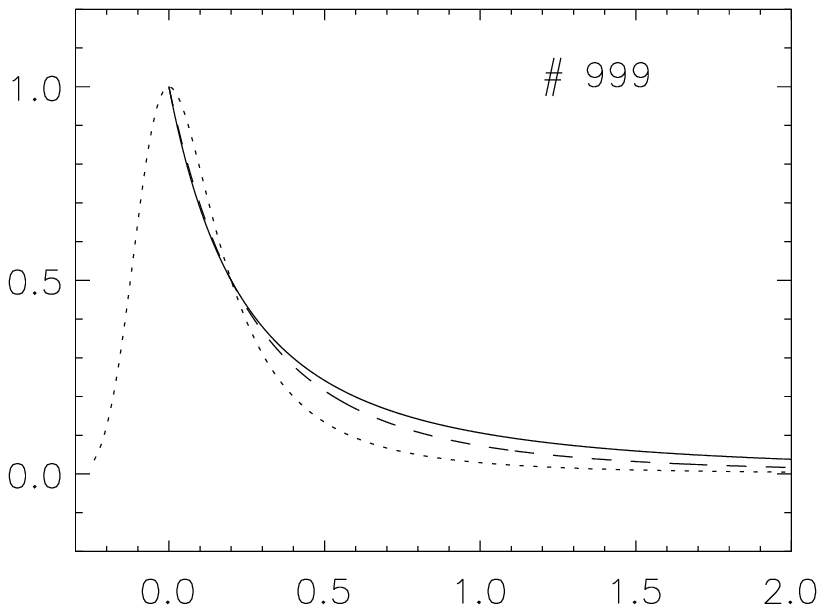}}
 \resizebox{5cm}{!}{\includegraphics{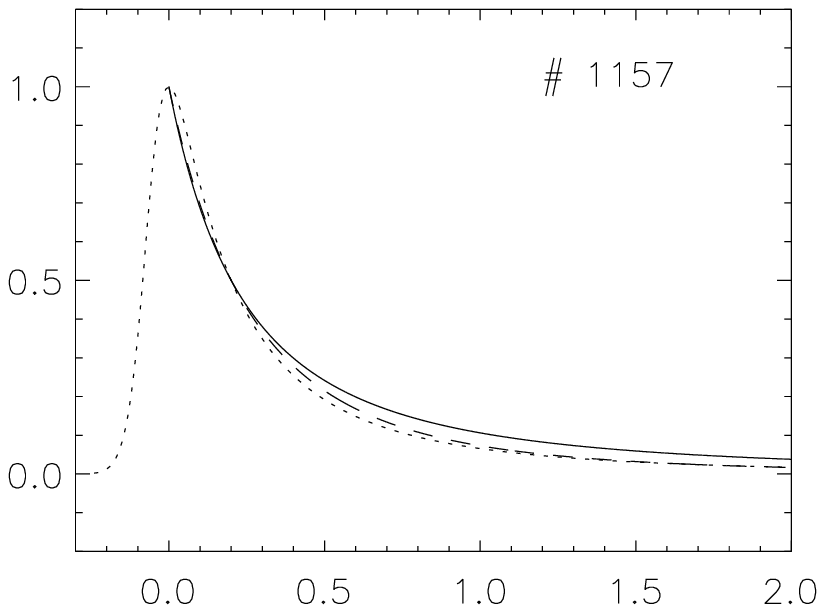}}
 \resizebox{5cm}{!}{\includegraphics{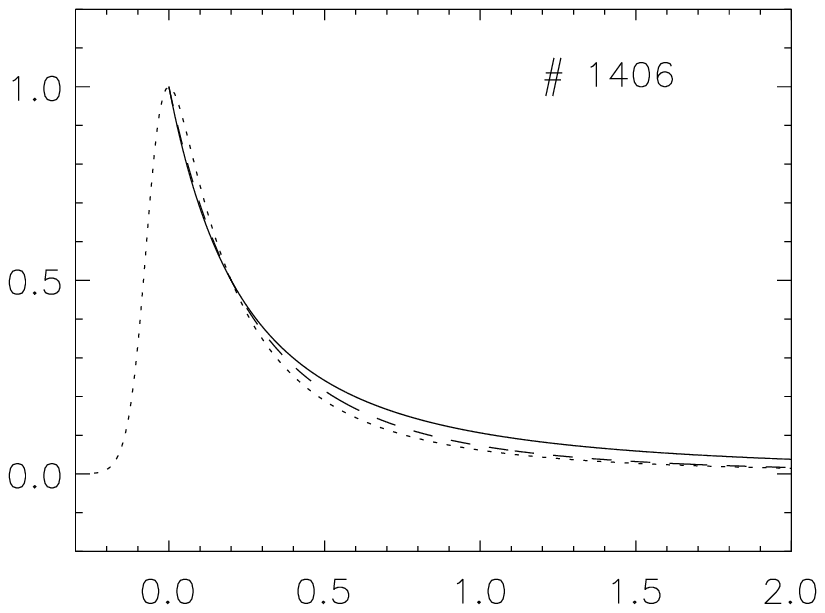}}
 \resizebox{5cm}{!}{\includegraphics{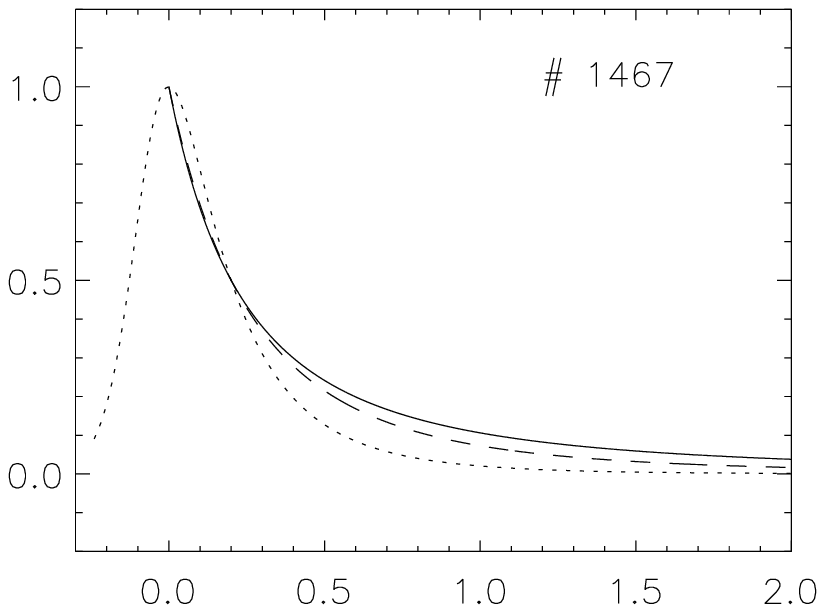}}
 \resizebox{5cm}{!}{\includegraphics{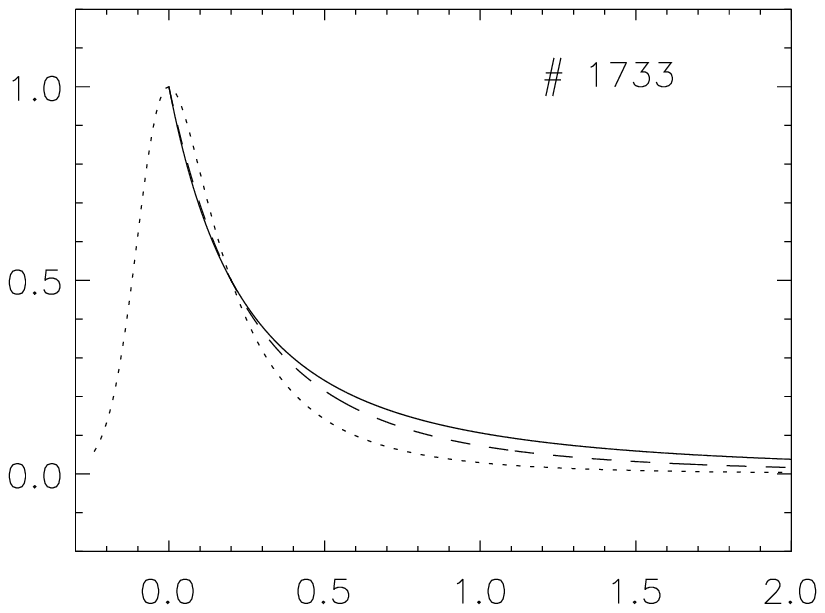}}
  \resizebox{5cm}{!}{\includegraphics{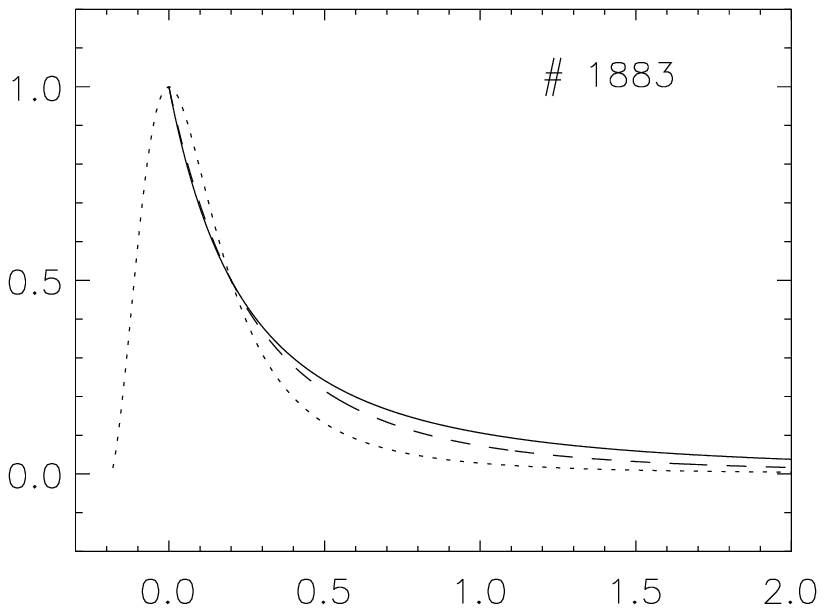}}
 \resizebox{5cm}{!}{\includegraphics{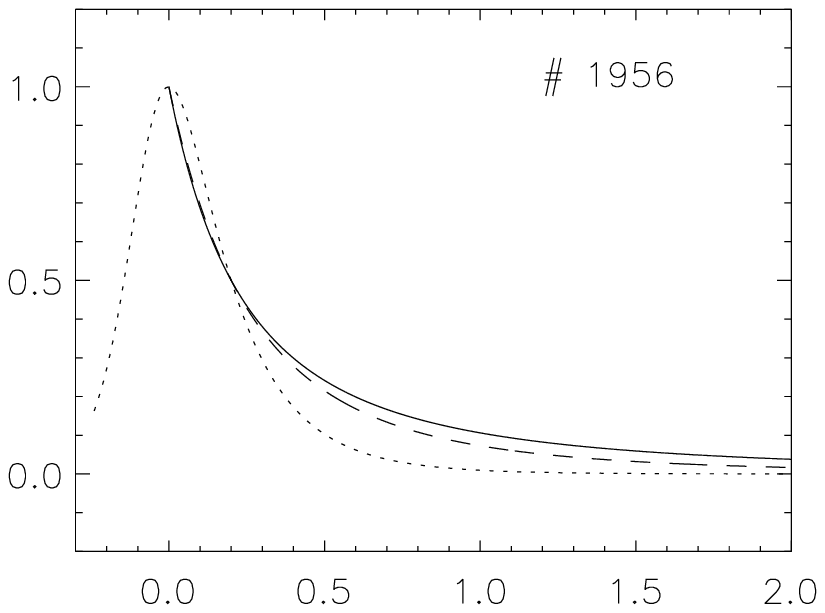}}
 \resizebox{5cm}{!}{\includegraphics{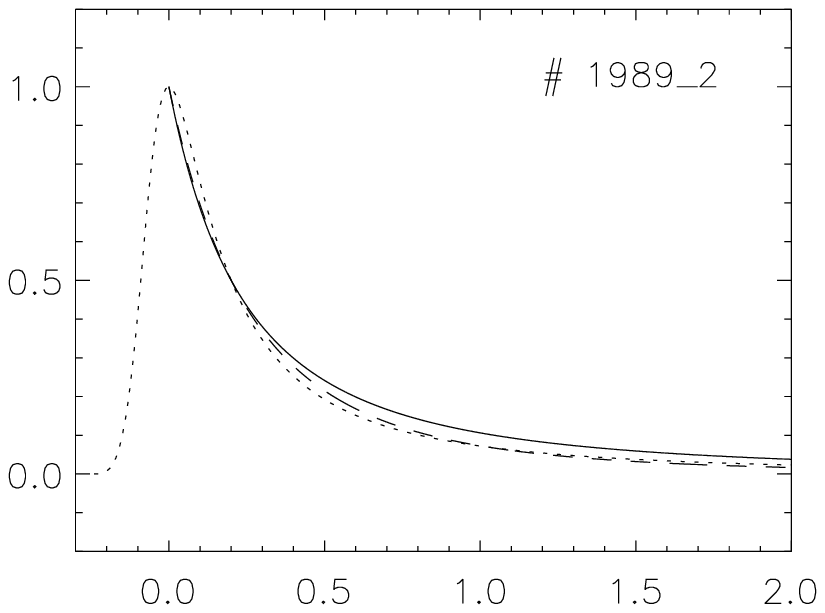}}
 \resizebox{5cm}{!}{\includegraphics{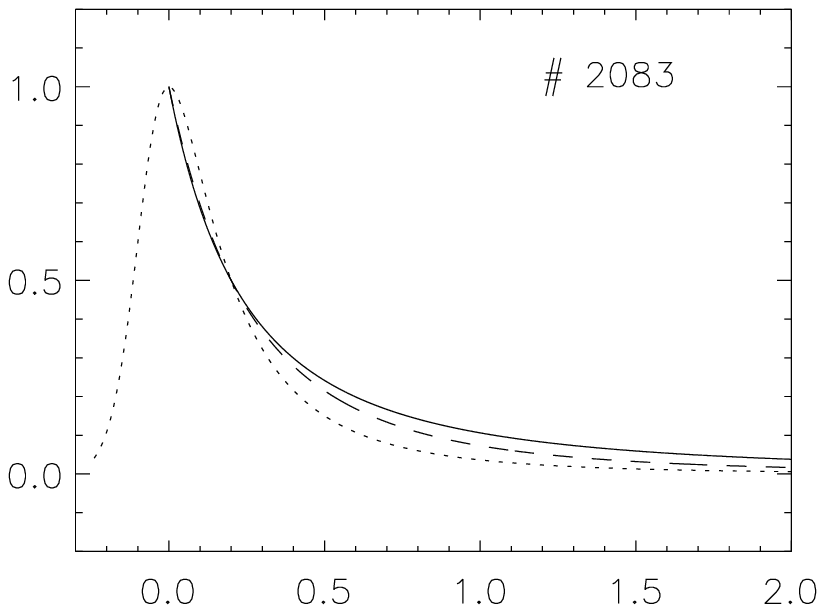}}
 \resizebox{5cm}{!}{\includegraphics{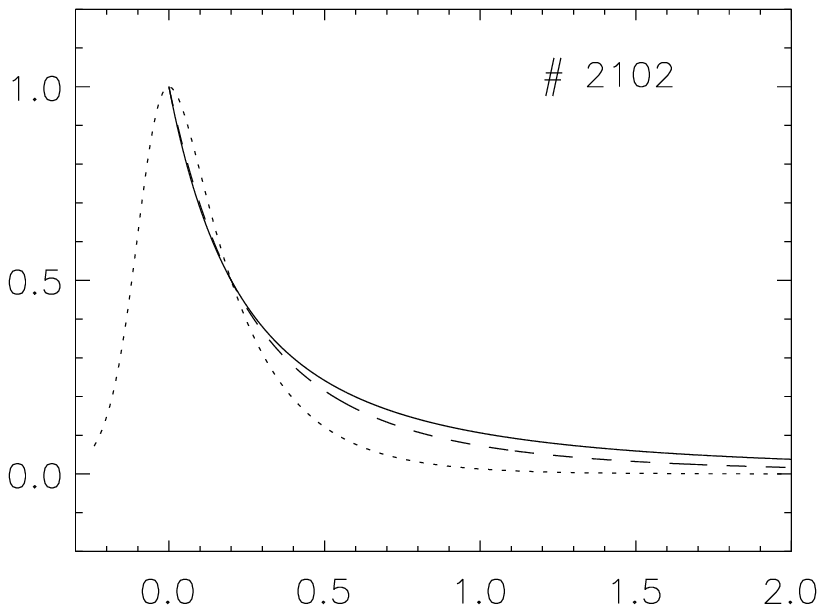}}
 \resizebox{5cm}{!}{\includegraphics{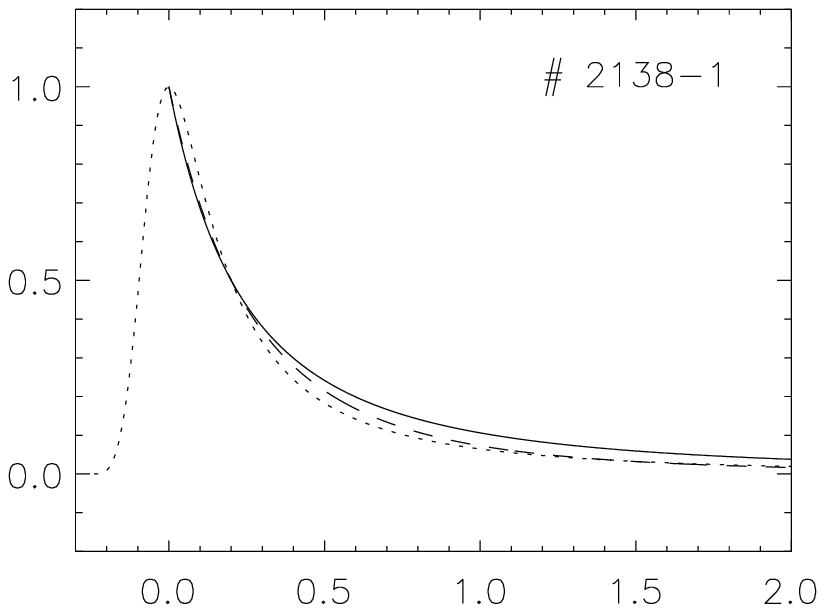}}
 \resizebox{5cm}{!}{\includegraphics{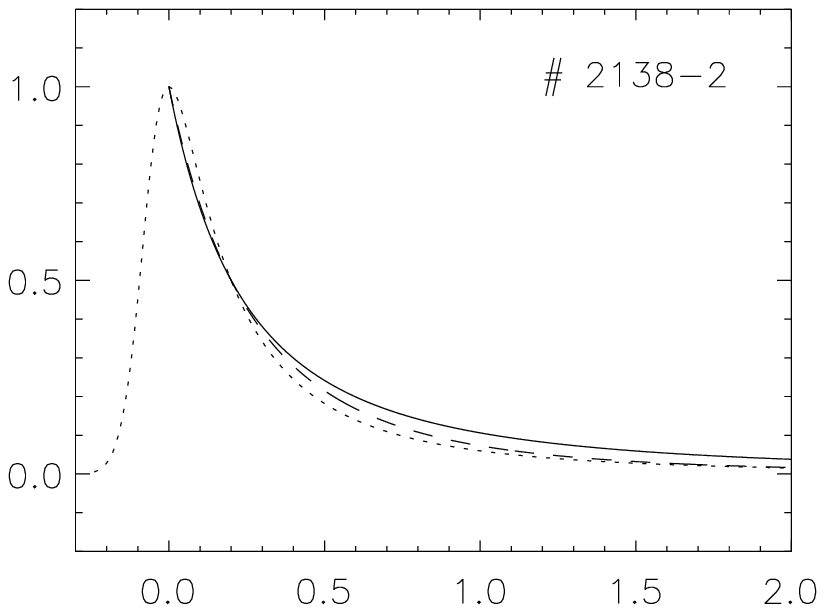}}
 \resizebox{5cm}{!}{\includegraphics{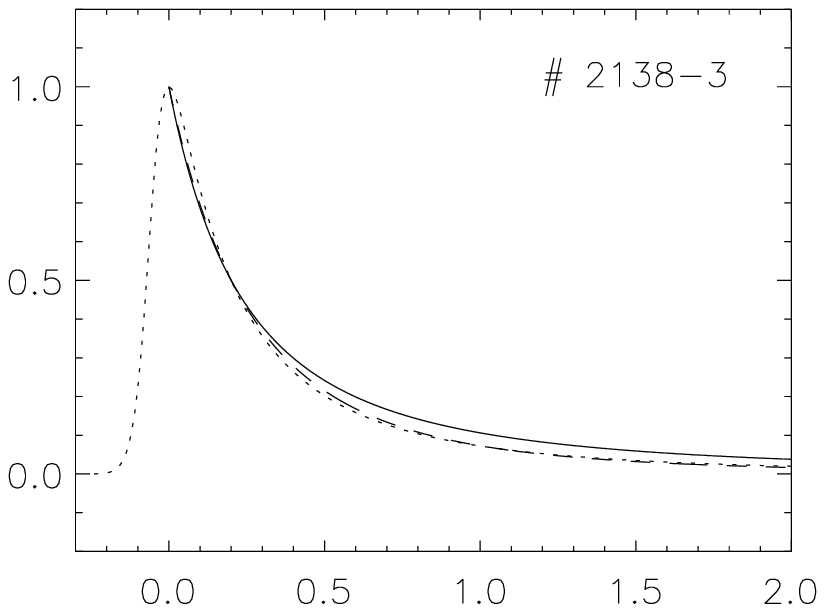}}

 \caption{--Plots of the fitting curves (the
dot line), the standard decay form (the dash line) and the
marginal decay curve (the solid line) for the 75 individual pulses
of the sample. }
 \label{sed}
\end{figure*}

\begin{figure}
\centering
\includegraphics[width=5in,angle=0]{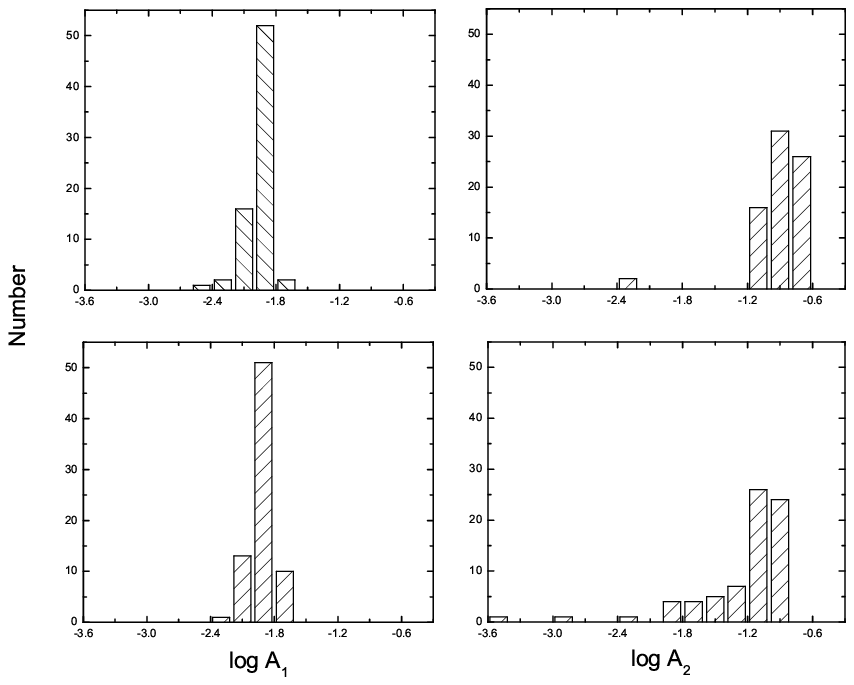}
\caption{ Distributions of $A_1$ and $A_2$ of all the pulses in
our sample. The two lower panels are associated with the marginal
decay curve and the two upper panels correspond to the standard
decay form (the two exclusive events are not included).}
\end{figure}

\section{discussion and conclusions  }

A reverse S-feature deviation of profiles of light curves arising
from local pulses containing a decaying tail from those associated
with very narrow or suddenly dimming local pulses, the so-called
standard decay form, could be seen in Fig. 3 of Paper II. We
investigate in this paper if FRED pulses observed bear indeed this
feature, and if they do how to measure the deviation. We define
two areas $A_1$ and $A_2$ to describe the deviations within and
beyond the $FWHM$ position in the decay phase, respectively.
Suggested in our analysis, different from the standard decay form,
there is a marginal decay curve which reflects the profile of the
light curve arising from a local $\delta$ function pulse with a
mono-color radiation. We employ a sufficiently large sample of
FRED pulses of GRBs to study this issue. The study shows that the
reverse S-feature indeed exists in the profiles of all the pulses
concerned when compared with the marginal decay curve, while 73
out of 75 individual pulses show the feature as well when compared
with the standard decay form. We also find that the values of
$A_1$ and $A_2$ for most pulses of the sample are quite large
which suggests that the corresponding local pulses might contain a
long decay time relative to the time scale of the curvature
effect.

For the two exclusive events, $\#$ 3257 and $\#$ 5495, the
deviation from the standard decay form is ``positive'', rather
than ``negative'', beyond the $FWHM$ position in the decay phase
(say, $\tau'>0.2$). We do not know what causes this exception.
Here are several outlets we can think of. The first is associated
with the background subtracting. We find that over or less
subtracting the background count would shift the corresponding
profile under or over the standard decay form beyond the $FWHM$
position in the decay phase. Illustrated in the left panel of Fig.
7 is this effect which is quite significant. The second is the
impact of the rest frame radiation form. Note that the standard
decay form is defined when the indexes of the rest frame spectrum
are taken as $\alpha_0=-1$ and $\beta_0=-2.25$. As shown in Preece
et al. (2000), the indexes could take many other values. We wonder
if different values of the indexes could lead to a much different
deviation. Shown in the right panel of Fig. 7 is this effect,
where two modified standard decay forms, for which the rest frame
spectral indexes of the standard decay form are replaced with
others, are presented. It indicates that different values of the
indexes could indeed change the profile beyond the $FWHM$ position
in the decay phase, but the deviation is relatively small (compare
the two panels of the figure).

\begin{figure}
\centering
\includegraphics[width=5in,angle=0]{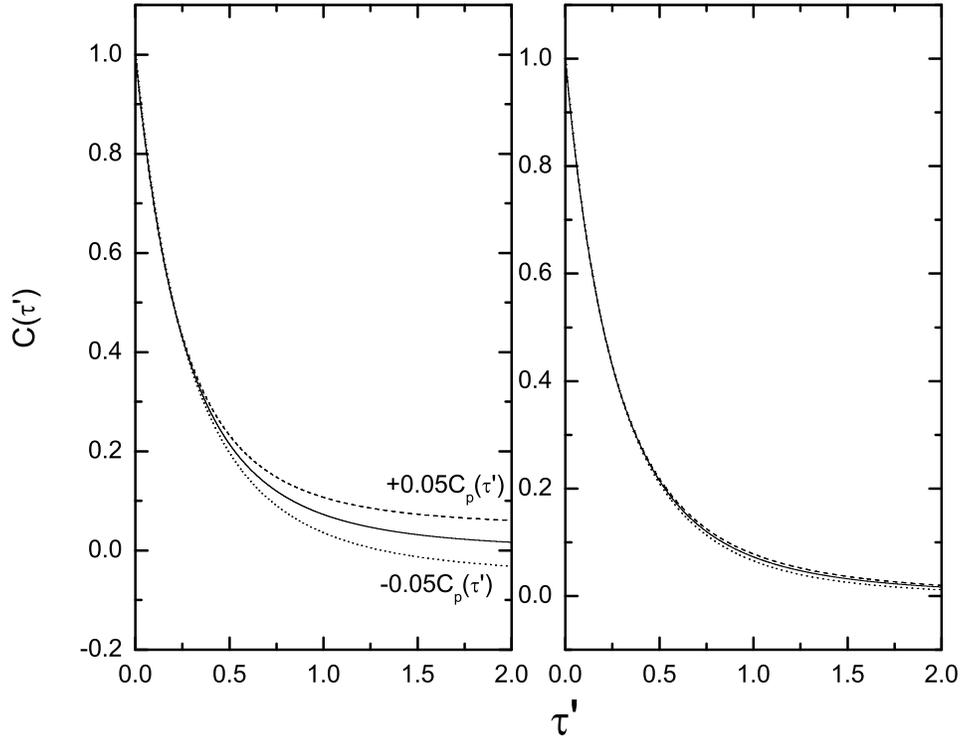}
\caption{The left panel is the plot of the effect of the
background subtracting, on the deviation from the standard decay
form, where the solid line is the standard decay form $C(\tau)$,
the dash line represents the curve of $C'(\tau)=C(\tau)+0.05C_{p}$
(which is associated with the case of less subtracting the
background count), and the dot line stands for the curve of
$C'(\tau)=C(\tau)-0.05C_{p}$ (which corresponds to the case of
over subtracting the background count). The right panel is the
plot showing the effect of the rest frame radiation form on the
deviation from the standard decay form, where the solid line is
the standard decay form (for which, $\alpha_0=-1$ and
$\beta_0=-2.25$ are adopted), the dash line is the modified
standard decay form when replacing $\beta_0=-2.25$ with
$\beta_0=-1$, and the dot line represents another modified
standard decay form by replacing $\alpha_0=-1$ with
$\alpha_0=-0.5$ and replacing $\beta_0=-2.25$ with
$\beta_0=-4.5$.}
\end{figure}

As suggested in Paper II, equation (1) holds in the case of
uniform jets. When $\theta _{\max }$ is very small (say, $\theta
_{\max }\sim 1/\Gamma$), there will be a turnover feature in the
decay tail of the light curve (see Paper II Fig. 2). One could
check that, when $\theta _{\max }$ is sufficiently large, the
turnover feature would not be detectable. In this case, a FRED
pulse would also be observed. Therefore, the conclusion above
holds as well in the case of uniform jets, as long as the opening
angle of jets is not extremely small.

Hinted by our analysis and the previous works (see Ryde and
Petrosian 2002; Paper II), one can conclude that the curvature
effect would lead to FRED pulses, and for the pulses caused by the
curvature effect their profiles would exhibit a reverse S-feature
deviation from the marginal decay curve. Thus we propose to take
the marginal decay curve as a criteria to check if an observed
pulse could be taken as a candidate suffered from the curvature
effect.

An interesting question arises accordingly, which is that, for
those non-FRED pulses of GRBs, what one could expect. We suspect
that, the reverse S-feature might no longer be observed in these
cases and such pulses might be associated with structure jets or
the scattering ejecta. This might deserve a further investigation
(see Lu and Qin 2005 in preparation).

We thank the anonymous referee who located some errors in the
original manuscript and made many helpful suggestions. This work
was supported by the Special Funds for Major State Basic Research
Projects (``973'') and National Natural Science Foundation of
China (No. 10273019 and No. 10463001).
\begin{table*}
\centering \caption{A list of the estimated values of the
deviation areas, $A_1$ and $A_2$, from the marginal decay curve
and the standard decay form, for the pulses of our sample}
\begin{tabular}{|c|c||c|c||c|c|c||c|c||c|}
\hline\hline  trigger number & $A_{1}(M)$  & $A_{2}(M)$  & $A_{1}$  & $A_{2}$ &trigger number & $A_{1}(M)$  & $A_{2}(M)$  & $A_{1}$  & $A_{2}$\\
\hline
   563 &    0.0139 &   -0.162 &    0.0126 &   -0.0976 &         3648-3 &    0.0127 &   -0.123 &    0.0115 &    -0.0578 \\
       907 &    0.0119 &   -0.0993 &    0.0107 &   -0.0344 &       3765 &    0.0160 &   -0.180 &    0.0147 &   -0.115 \\
       914 &    0.0122 &   -0.108 &    0.0109 &   -0.0432 &       3870 &    0.00853 &   -0.0915 &    0.00723 &   -0.0266 \\
     973-1 &    0.0139 &   -0.140 &     0.0126 &   -0.0749 &       3875 &    0.00844 &    -0.136 &    0.00714 &   -0.0718 \\
     973-2 &     0.0133 &   -0.138 &    0.0120 &   -0.0735 &       3886 &    0.0133 &   -0.137 &    0.0120 &   -0.0723 \\
       999 &    0.0142 &   -0.146 &    0.0129 &   -0.0809 &       3892 &      0.014 &   -0.141 &     0.0127 &   -0.0766 \\
      1157 &    0.00941 &   -0.0786 &    0.00811 &   -0.0137 &       3954 &    0.0134 &   -0.144&    0.0121 &   -0.0792 \\
      1406 &     0.00920 &   -0.0863 &     0.0079 &    -0.0214 &       4157 &    0.0125 &   -0.146 &    0.0112 &    -0.0811 \\
      1467 &    0.0140 &   -0.159 &    0.0127 &   -0.0936 &     4350-1 &    0.0168 &   -0.192 &    0.0155 &   -0.127 \\
      1733 &    0.0132 &    -0.144 &    0.0119 &   -0.0790 &     4350-2 &    0.00665 &    -0.135 &    0.00535 &   -0.0705 \\
      1883 &    0.0141 &   -0.148 &    0.0128 &   -0.0829 &     4350-3 &    0.0160 &    -0.186 &    0.0147 &   -0.122 \\
      1956 &    0.0154 &   -0.178 &    0.0141 &    -0.113 &       4368 &    0.0131 &   -0.128 &    0.0118 &   -0.0634 \\
      1989 &    0.0103 &   -0.0653 &    0.00896 &   -3.98E-4 &     5478-1 &    0.0153 &   -0.173 &    0.0140 &   -0.108 \\
      2083 &    0.0129 &   -0.132 &    0.0116 &   -0.0675 &     5478-2 &    0.0180 &   -0.201 &    0.0167 &   -0.136 \\
      2102 &    0.0135 &   -0.167 &    0.0122 &   -0.103 &       5495 &    0.00704 &   -0.00433 &    0.00574 &    0.0605 \\
    2138-1 &    0.0111 &   -0.0800 &    0.00976 &   -0.0151 &       5517 &    0.0143 &   -0.152 &    0.0130 &   -0.0878 \\
    2138-2 &    0.0106 &   -0.0895 &    0.00934 &   -0.0246 &       5523 &    0.0135 &   -0.171 &    0.0123 &   -0.107 \\
    2138-3 &    0.00789 &   -0.0660 &    0.00659 &   -0.00118 &       5541 &    0.0115 &   -0.112 &    0.0103 &    -0.0471 \\
      2193 &    0.0153 &   -0.180 &    0.0139 &   -0.115 &       5601 &     0.0136 &   -0.133 &     0.0123 &   -0.0690 \\
      2387 &    0.0165 &   -0.193 &    0.0152 &   -0.128 &       6159 &    0.0117 &   -0.149 &    0.0105 &   -0.0849 \\
      2484 &    0.0168 &   -0.194 &    0.0155 &   -0.129 &       6335 &     0.0118 &    -0.105 &     0.0105 &   -0.0406 \\
      2519 &    0.00525 &   -0.0845 &    0.00395 &   -0.0196 &       6397 &    0.0123 &   -0.106 &    0.0110 &   -0.0421 \\
      2530 &    0.0139 &   -0.173 &    0.0125 &   -0.108 &       6504 &    0.0158 &   -0.185 &    0.0145 &   -0.120 \\
      2662 &     0.00850 &   -0.134 &     0.0072 &   -0.0689 &       6621 &     0.0104 &   -0.0755 &     0.0091 &   -0.0107 \\
      2665 &     0.0117 &   -0.0949 &     0.0104 &   -0.0300 &       6625 &    0.0130 &   -0.168 &    0.0117 &    -0.104 \\
      2700 &    0.0130 &   -0.168 &    0.0117 &   -0.103 &       6672 &    0.0155 &   -0.170 &    0.0142 &   -0.106 \\
      2880 &    0.0129 &   -0.120 &    0.0115 &   -0.0547 &       6930 &    0.0178 &   -0.201 &    0.0165 &   -0.137 \\
      2919 &    0.0108 &   -0.09725 &    0.00945 &   -0.0323 &       7293 &    0.0117 &   -0.0937 &    0.0104 &   -0.0289 \\
      3003 &    0.0138 &    -0.142 &    0.0125 &   -0.0770 &       7295 &    0.00798 &   -0.121&    0.00668 &   -0.0565 \\
      3143 &    0.0125 &   -0.136 &    0.0112 &   -0.0709 &       7475 &    0.0157 &   -0.174 &    0.0144 &   -0.109 \\
      3155 &    0.0144 &   -0.177 &    0.0131 &   -0.112 &       7548 &    0.0116 &   -0.0896 &    0.0103 &   -0.0248 \\
      3256 &    0.0153 &   -0.192 &    0.0140 &   -0.126&       7588 &    0.0142 &   -0.168 &    0.0129 &    -0.103 \\
      3257 &    0.00816 &   -0.00482 &    0.00686 &    0.0683 &       7638 &    0.00709 &   -0.0704 &    0.00579 &   -0.00559 \\
      3290 &    0.00956 &   -0.0786 &    0.00826 &   -0.0137 &       7648 &    0.0167 &   -0.193 &    0.0154 &   -0.129 \\
    3415-1 &    0.00886 &   -0.130 &    0.00756 &   -0.0648 &       7711 &    0.0141 &   -0.152 &    0.0128 &   -0.0877 \\
    3415-2 &    0.0108 &   -0.157 &    0.00954 &   -0.0916 &       8049 &    0.0166 &   -0.185 &    0.0154 &   -0.121 \\
    3648-1 &    0.0138 &   -0.148 &    0.0125 &   -0.0827 &       8111 &    0.0105 &   -0.150 &    0.00926 &   -0.0855 \\
    3648-2 &    0.0144 &   -0.168 &    0.0130 &    -0.1032 &            &            &            &            &            \\
 \hline
\end{tabular}

Note: $A_{1}(M)$ and $A_{2}(M)$ are the two deviations of the
profile of a light curve from the marginal decay curve.
\end{table*}

\label{lastpage}


\clearpage
\begin{thebibliography}{10}
\bibitem[Band et al. (1993)]{Ba93}  Band, D.,Matteson, J., Ford, L., Schaefer, B., Palmer, D.,
 Teegarden, B., Cline, T., Briggs, M., et al. 1993, ApJ, 413, 281

\bibitem[Fenimore et al. (1996)]{Fe96}  Fenimore, E. E., Madras, C. D., and
Nayakshin, S. 1996, ApJ, 473, 998
\bibitem[Fishman et al. (1994)]{Fi94}  Fishman, G. J.,Gerald J., Meegan, Charles A., Wilson, Robert B.,
 Brock, Martin N., Horack, John M., Kouveliotou, Chryssa, Howard, Sethanne, et al. 1994, ApJS, 92, 229
\bibitem[Fishman et al. (1995)]{Fi95} Fishman, G., Meegan, C. 1995, ARA\&A, 33, 415
\bibitem[Ford et al. (1995)]{Fo95}  Ford, L. A., Band, D. L., Matteson, J. L., Briggs, M. S.,
 Pendleton, G. N., Preece, R. D., Paciesas, W. S., et al. 1995, ApJ, 439, 307


\bibitem[Friedman et al. (2001)]{Pa01} Friedman, Andrew S. and Bloom, Joshua
S. 2005, APJ, 627, 1F
\bibitem[Fruchter et al. (1999)]{Fr99}  Fruchter, A. S., Thorsett, S. E., Metzger, Mark R., Sahu, Kailash C.,
 Petro, Larry, Livio, Mario, Ferguson, Henry, Pian, Elena, et al. 1999, APJ, 519,L13

\bibitem[Kocevski et al. (2003)]{Ko03}  Kocevski, D., Ryde, F., and Liang,
E. 2003, ApJ, 596, 389 (Paper I)

\bibitem[Lee et al.(2005)]{Le00} Lee, A., Bloom, E. D., \& Petrosian, V. 2000a, ApJS, 131,1
\bibitem[Lee et al. (2005)]{Le00} Lee, A., Bloom, E. D., \& Petrosian, V. 2000b, ApJS, 131, 21

\bibitem[Norris et al. (1996)]{No96}  Norris, J. P., Nemiroff, R. J., Bonnell, J. T.,
 Scargle, J. D., Kouveliotou, C., Paciesas, W. S., Meegan, C. A. and Fishman, G. J. 1996, ApJ, 459, 393

\bibitem[Piran (1999)]{Pi99}Piran, T. 1999, Phys. Rep., 314, 575
\bibitem[Preece et al. (1998)]{Pr98}  Preece, R. D.,Pendleton, Geoffrey N., Briggs, Michael S.,
 Mallozzi, Robert S., Paciesas, William S., Band, David L., Matteson, James L., Meegan, C. A.  1998, ApJ, 496, 849
\bibitem[Preece et al. (2000)]{Pr00}  Preece, R. D., Briggs, M. S., Mallozzi, R. S., Pendleton, G. N.,
 Paciesas, W. S. and Band, D. L. 2000, ApJS, 126, 19

\bibitem[Qin (2002)]{Qi02}  Qin, Y.-P. 2002, A\&A, 396, 705
\bibitem[Qin (2003)]{Qi03}  Qin, Y.-P. 2003, A\&A, 407, 393
\bibitem[Qin et al. 2004]{Qi04}Qin, Y. P., Zhang Z. B., Zhang F. W. and Cui X. H. 2004, APJ,
617, 439 (Paper II)

\bibitem[Ryde and Svensson (2002)]{Ry00}   Ryde, F., \& Svensson, R. 2000, ApJ, 529, L13
\bibitem[Ryde and Petrosian (2002)]{Ry02}  Ryde, F., and Petrosian, V. 2002,
ApJ, 578, 290

\bibitem[Schaefer et al. (1994)]{Sc94}  Schaefer, B. E., Teegarden, Bonnard J., Fantasia, Stephan F.,
 Palmer, David, Cline, Thomas L., Matteson, James L., Band, David L., Ford, Lyle A., et al. 1994, ApJS, 92, 285


\end{thebibliography}
\end{document}